\documentclass{article}

\usepackage{times}
\usepackage{graphicx} 
\usepackage{subfigure} 

\usepackage{natbib}
\usepackage{algorithm}
\usepackage{algorithmic}

\usepackage{amsmath}
\usepackage{amssymb}
\usepackage{mathrsfs}
\usepackage{hyperref}
\usepackage{bm}
\usepackage{multirow}
\usepackage{graphicx}
\usepackage{caption}
\usepackage{array}
\usepackage{tabu}
\usepackage{hyperref}

\usepackage[accepted]{icml2016}

\usepackage[utf8]{inputenc}
\usepackage[english]{babel}

\icmltitlerunning{Invertible Autoencoder for domain adaptation}

\begin{document} 

\twocolumn[
\icmltitle{Invertible Autoencoder for domain adaptation}

\icmlauthor{Yunfei Teng}{yt1208@nyu.edu }
\icmlauthor{Anna Choromanska}{ac5455@nyu.edu}
\icmlauthor{Mariusz Bojarski}{mbojarski@nvidia.com}

\icmlkeywords{}
\vskip 0.3in
]

\begin{abstract} 
The unsupervised image-to-image translation aims at finding a mapping between the source ($\mathcal{A}$) and target ($\mathcal{B}$) image domains, where in many applications aligned image pairs are not available at training. This is an ill-posed learning problem since it requires inferring the joint probability distribution from marginals. Joint learning of coupled mappings $\mathcal{F}_{\mathcal{A}\mathcal{B}}: \mathcal{A} \rightarrow \mathcal{B}$ and $\mathcal{F}_{\mathcal{B}\mathcal{A}}: \mathcal{B} \rightarrow \mathcal{A}$ is commonly used by the state-of-the-art methods, like CycleGAN~\cite{CycleGAN2017}, to learn this translation by introducing cycle consistency requirement to the learning problem, i.e. $\mathcal{F}_{\mathcal{A}\mathcal{B}}(\mathcal{F}_{\mathcal{B}\mathcal{A}}(\mathcal{B})) \approx \mathcal{B}$ and $\mathcal{F}_{\mathcal{B}\mathcal{A}}(\mathcal{F}_{\mathcal{A}\mathcal{B}}(\mathcal{A})) \approx \mathcal{A}$. Cycle consistency enforces the preservation of the mutual information between input and translated images. However, it does not explicitly enforce $\mathcal{F}_{\mathcal{B}\mathcal{A}}$ to be an inverse operation to $\mathcal{F}_{\mathcal{A}\mathcal{B}}$. We propose a new deep architecture that we call \textit{invertible autoencoder (InvAuto)} to explicitly enforce this relation. This is done by forcing an encoder to be an inverted version of the decoder, where corresponding layers perform opposite mappings and share parameters. The mappings are constrained to be orthonormal. The resulting architecture leads to the reduction of the number of trainable parameters (up to $2$ times). We present image translation results on benchmark data sets and demonstrate state-of-the art performance of our approach. Finally, we test the proposed domain adaptation method on the task of road video conversion. We demonstrate that the videos converted with InvAuto have high quality and show that the NVIDIA neural-network-based end-to-end learning system for autonomous driving, known as PilotNet, trained on real road videos performs well when tested on the converted ones.
\end{abstract} 

\section{Introduction}
\label{Introduction}

\begin{figure*}[h!]
\centering
  \begin{table}[H]
  \begin{tabu} to \textwidth {X[c] X[c] X[c] X[c]}
  \large{\textbf{InvAuto}} &\large{\textbf{Auto}} &\large{\textbf{Cycle}} &\large{\textbf{VAE}}
  \end{tabu}
  \end{table}
  \vspace{-0.2in}
  \subfigure[MNIST MLP]{\includegraphics[width=0.24\textwidth]{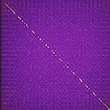} \label{1_1}}
  \subfigure[MNIST MLP]{\includegraphics[width=0.24\textwidth]{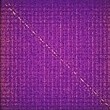} \label{1_2}}
  \subfigure[MNIST MLP]{\includegraphics[width=0.24\textwidth]{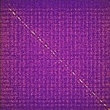} \label{1_3}}
  \subfigure[MNIST MLP]{\includegraphics[width=0.24\textwidth]{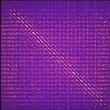} \label{1_4}}
  \subfigure[CIFAR ResNet]{\includegraphics[width=0.24\textwidth]{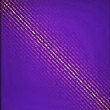} \label{4_1}}
  \subfigure[CIFAR ResNet]{\includegraphics[width=0.24\textwidth]{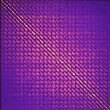} \label{4_2}}
  \subfigure[CIFAR ResNet]{\includegraphics[width=0.24\textwidth]{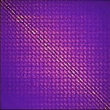} \label{4_3}}
  \subfigure[CIFAR ResNet]{\includegraphics[width=0.24\textwidth]{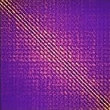} \label{4_4}}
 \vspace{-0.1in}
\caption{Heatmap of the values of matrix $DE$ for InvAuto (a and e), Auto (b and f), Cycle (c and g), and VAE (d and h) on MLP and ResNet architectures and MNIST and CIFAR data sets. Matrices $E$ and $D$ are constructed by multiplying the weight matrices of consecutive layers of encoder and decoder, respectively. In case of InvAuto, $DE$ is the closest to the identity matrix.}
\label{fig:examplea}
 \vspace{-0.05in}
\end{figure*}

Inter-domain translation problem of converting an instance, e.g.: image or video, from one domain to another is applicable to a wide variety of learning tasks, including object detection and recognition, image categorization, sentiment analysis, action recognition, speech recognition, and more. High-quality domain translators ensure that an arbitrary learning model trained on the samples from the source domain, can perform well when tested on the translated samples\footnote{Similarly, an arbitrary learning model trained on the translated samples should perform well on the samples from the target domain. Training in this framework is however much more computationally expensive.}. The translation problem can be posed in the supervised learning framework, e.g.:~\cite{pix2pix, pix2pixhd}, where the learner has access to corresponding pairs of instances from both domains, or unsupervised learning framework, e.g.:~\cite{CycleGAN2017,unit}, where no such paired instances are available. This paper focuses on the latter case, which is more difficult but at the same time more realistic as acquiring the data set of paired images is often impossible in practice.

The unsupervised domain adaptation is typically solved using generative adversarial networks (GAN) framework~\cite{gan_ian}, where the generator performs domain translation and is trained to learn the mapping from the source to the target domain and the discriminator is trained to discriminate between original images from the target domain and those provided by the generator. In this setting, the generator usually has the structure of the autoencoder. The two most common state-of-the-art domain adaptation approaches, CycleGAN~\cite{CycleGAN2017} and UNIT~\cite{unit}, are built on this basic approach. CycleGAN addresses the problem of adaptation from domain $\mathcal{A}$ to domain $\mathcal{B}$ by training two translation networks, where one realizes the mapping $\mathcal{F}_{\mathcal{A}\mathcal{B}}$ and the other realizes $\mathcal{F}_{\mathcal{B}\mathcal{A}}$. The cycle consistency loss ensures the correlation between input image and the corresponding translation. In particular, to achieve cycle consistency, CycleGAN trains two autoencoders, where each minimizes its own adversarial loss and they both jointly minimize 

\vspace{-0.25in}
\begin{equation}
\|\mathcal{F}_{\mathcal{A}\mathcal{B}}(\mathcal{F}_{\mathcal{B}\mathcal{A}}(\mathcal{B})) - \mathcal{B}\|_2^2 \:\:\:\text{and}\:\:\:\|\mathcal{F}_{\mathcal{B}\mathcal{A}}(\mathcal{F}_{\mathcal{A}\mathcal{B}}(\mathcal{A})) - \mathcal{A}\|_2^2.
\end{equation}

\vspace{-0.1in}
Cycle consistency loss is also incorporated into the recent implementations of UNIT. It is implicitly assumed that the model will learn the mappings $\mathcal{F}_{\mathcal{A}\mathcal{B}}$ and $\mathcal{F}_{\mathcal{B}\mathcal{A}}$ in such a way that $\mathcal{F}_{\mathcal{A}\mathcal{B}} = \mathcal{F}_{\mathcal{B}\mathcal{A}}^{-1}$, however it is not explicitly imposed. Consider a simple example. Assume the first autonecoder is a $2$-layer linear multi-layer perceptron (MLP) where the  weight matrix of the first layer (encoder) is denoted as $E_1$ and the weight matrix of the second layer (decoder) is denoted as $D_1$.  Thus, for an input $x_{\mathcal{A}} \!\in\! \mathcal{A}$ it outputs $y_{\mathcal{B}}(x_{\mathcal{A}}) \!=\! D_1E_1x_{\mathcal{A}}$. The second autoencoder then is a $2$-layer MLP with encoder weight matrix $E_2$ and decoder weight matrix $D_2$ that for an input data point $x_{\mathcal{B}}$ should produce output $y_{\mathcal{A}}(x_{\mathcal{B}}) \!=\! D_2E_2x_{\mathcal{B}}$. To satisfy cycle consistency requirement, the following should hold: $y_{\mathcal{A}}(y_{\mathcal{B}}(x_{\mathcal{A}})) \!=\! (x_{\mathcal{A}})$ and $y_{\mathcal{B}}(y_{\mathcal{A}}(x_{\mathcal{B}})) \!=\! (x_{\mathcal{B}})$. These two conditions are equivalent to $D_2E_2D_1E_1 \!=\! I$ and $D_1E_1D_2E_2 \!=\! I$. This holds for example when $D_1 \!=\! E_2^{-1}$ and $D_2 \!=\! E_1^{-1}$.  

In contrast to this approach, we implicitly require $\mathcal{F}_{\mathcal{A}\mathcal{B}} = \mathcal{F}_{\mathcal{B}\mathcal{A}}^{-1}$. Thus, in the context of the given simple example, we correlate encoders and decoders to satisfy inversion conditions $D_1 = E_2^{-1}$ and $D_2 = E_1^{-1}$. We avoid performing prohibitive inversions of large matrices and instead guarantee these conditions to hold through two steps: (i) introducing shared parametrization of encoder $E_2$ and decoder $D_1$ such that $D_1 = E_2^{\top}$ ($E_1$ and $D_2$ is treated similarly) and (ii) appropriate training to achieve orthonormality $E_2^{\top} = E_2^{-1}$ and $E_1^{\top} = E_1^{-1}$, i.e. we train autoencoder $(E_2,D_1)$ to satisfy $D_1E_2x_{\mathcal{B}} = x_{\mathcal{B}}$ for arbitrary input $x_{\mathcal{B}}$ and autoencoder $(E_1,D_2)$ to satisfy $D_2E_1x_{\mathcal{A}} = x_{\mathcal{A}}$ for arbitrary input $x_{\mathcal{A}}$. Since the encoder and decoder are coupled as given in (i), such training leads to satisfying inversion conditions. Practical networks contain linear and non-linear transformations. We therefore propose specific architectures, which are invertible. 

\begin{figure}[h!]
\vspace{-0.05in}
\centering
\begin{minipage}[c]{\textwidth}
\hspace{-0.1in}\includegraphics[width=0.5 \textwidth]{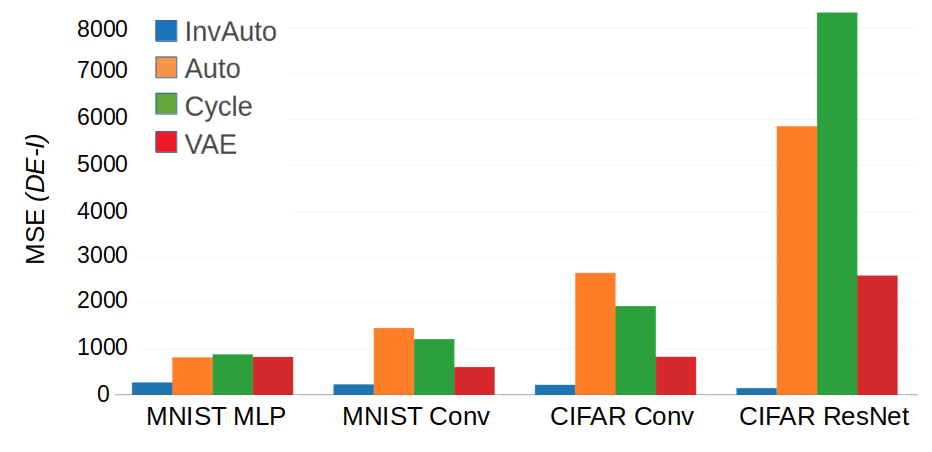}
\end{minipage}
\vspace{-0.22in}
\caption{Comparison of the mean squared error (MSE) $\text{MSE}(DE - I)$ for InvAuto, Auto, Cycle, and VAE on MLP, convolutional, and ResNet architectures and MNIST and CIFAR data sets. Matrices $E$ and $D$ are constructed by multiplying the weight matrices of consecutive layers of encoder and decoder, respectively.}
\label{fig:exampleb}
\vspace{-0.05in}
\end{figure}

Figure~\ref{fig:examplea} (see also its extended version, Figure~\ref{fig:exampleasup}, in the Supplement) and~\ref{fig:exampleb} illustrate the basic idea behind InvAuto. The plots were obtained by training a single autoencoder $(E,D)$ to reconstruct its input. InvAuto has shared weights satisfying $D = E^{\top}$ and inverted non-linearities and clearly obtains matrix $DE$ that is the closest to identity compared to other methods, i.e. vanilla autoencoder (Auto), autoencoder with cycle consistency (Cycle), and variational autoencoder (VAE)~\cite{vae}. Note also that at the same time InvAuto requires half of the number of trainable parameters.

This paper is organized as follows: Section~\ref{sec:rw} reviews the literature, Section~\ref{sec:ia} explains InvAuto in details, Section~\ref{sec:iada} explains how to apply InvAuto to domain adaptation, Section~\ref{Experiments} demonstrates experimental verification of the proposed approach, and Section~\ref{sec:c} provides conclusions.

\section{Related Work} 
\label{sec:rw}

Unsupervised image-to-image translation models were developed to tackle domain adaptation problem with unpaired data sets. A plethora of existing approaches utilize autoencoders trained in the GAN framework, where autoencoder serves as a generator, for this learning problem. This includes approaches based on conditional GAN~\cite{dong_2017,pix2pixhd} and methods introducing additional components to the loss function forcing partial cycle consistency~\cite{fbgan}. Another approach~\cite{cogan} introduces two coupled GANs, where each generator is an autoencoder and the coupling is obtained by sharing a subset of weights between autoencoders as well as between discriminators. This technique was later on extended to utilize variational autoencoders as generators~\cite{unit}. The resulting approach is commonly known as UNIT. CycleGAN presents yet another way of addressing the image-to-image translation by specific training scheme that preserves the mutual information between input and translated images~\cite{vincent_larochelle_bengio_manzagol_2008}. Both UNIT and CycleGAN constitute the most popular choices for performing image-to-image translation.

There also exist other learning tasks that can be viewed as instances of image-to-image translation problem. Among them, notable approaches focus on style transfer~\cite{gatys_2016, styletrans_2016, texture, color}. They aim at preserving the content of the input image while altering its style to mimic the style of the images from the target domain.
This goal is achieved by introducing content and style loss functions that are jointly optimized. Finally, inverse problems, such as super-resolution, also fall into the category of image-to-image translation problems~\cite{DBLP:journals/spm/McCannJU17}.

\section{Invertible autoencoder}
\label{sec:ia}

Here we explain the details of the architecture of InvAuto. The architecture needs to be symmetric to allow invertibility, e.g.: the layers should be arranged as $(\underbrace{T_1,T_2,\dots,T_M}_{\text{encoder $E$}},\underbrace{T_M^{-1},T_{M-1}^{-1},\dots,T_1^{-1})}_{\text{decoder $D$}}$, where $T_1,T_2,\dots,T_M$ denote subsequent transformations of the signal that is being propagated through the network ($M$ is the total number of those) and $T_1^{-1},T_2^{-1},\dots,T_M^{-1}$ denote their inversions. Thus, the architecture is inverted layer by layer, where any layer of the encoder has its mirror inverted counterpart in the decoder. The autoencoder is trained to reconstruct its input. Below we explain how to invert different types of layers of the deep model. 

\subsection{Fully-connected layer}

Consider transformation $T^E$ of an input signal performed by an arbitrary fully-connected layer of an encoder $E$ parametrized with weight matrix $W$. Let $x$ denote layer's input and $y$ denote its output. Thus 

\vspace{-0.1in}
\begin{equation}
T^E: y = Wx.
\label{eq:a}
\end{equation}
An inverse operation is then defined as
\begin{equation}
(T^E)^{-1}: x = W^{-1}y,
\label{eq:b}
\end{equation}
We parametrize the counterpart layer of the decoder with a transpose of $W$, thus the considered encoder and decoder layers will share parametrization. Therefore, we enforce the counterpart decoder's layer to perform transformation:
\begin{equation}
T^{D}: x = W^{\top}y.
\label{eq:c}
\end{equation}
By training the autoencoder to reconstruct its input on its output we will enforce orthonormality $W^{-1} = W^{\top}$ and thus equivalence of transformations $(T^E)^{-1}$ and $T^D$, i.e. $(T^E)^{-1} \equiv T^D$.

\subsection{Convolutional layer} 

Consider transformation $T^E$ of an input image performed by an arbitrary convolutional layer of an encoder $E$. Let $x$ denote layer's vectorized input image and $y$ denote corresponding output. $2$D convolution can be implemented using matrix multiplication involving a Toeplitz matrix~\cite{DBLP:conf/asap/VasudevanAG17}. Toeplitz matrix is obtained from the set of kernels of the $2$D convolutional filters . Thus transformation $T^E$ and its inverse $(T^E)^{-1}$ can be explained with the same equations as the ones used before, Equations~\ref{eq:a} and~\ref{eq:b}, however now $W$ is a Toeplitz matrix. We will again parametrize the counterpart layer of the decoder with a transpose of a Toeplitz matrix $W$. The transpose of the Toeplitz matrix is in practice obtained by copying weights from the considered convolutional layer to the counterpart decoder's layer that is implemented as a transposed convolutional layer (also known as a deconvolutional layer). Therefore, as before, we enforce the counterpart decoder's layer to perform transformation $T^{D}: x = W^{\top}y$ and by appropriate training ensure $(T^E)^{-1} \equiv T^D$.

\subsection{Activation function}

Invertible activation function should be a bijection. In this paper we consider a modified LeakyReLU activation function $\sigma$ and use only this non-linearity in the model. Consider transformation $T^E$ of an input signal performed by this non-linearity applied in the encoder $E$. This non-linearity is defined as

\vspace{-0.2in}
\begin{equation}\label{eq:for_act}
T^E: y = \sigma(x)= 
\begin{cases}
    \frac{1}{\alpha}x,  & \textit{if  } x\geq 0\\
    \alpha x,           & \textit{otherwise}.
\end{cases}    
\end{equation} 
\vspace{-0.1in}

An inverse operation is then defined as

\vspace{-0.15in}
\begin{equation}
(T^E)^{-1}: x = \sigma^{-1}(y)= 
\begin{cases}
    \alpha y,             & \textit{if  } x\geq 0\\
    \frac{1}{\alpha} y,   & \textit{otherwise}.
\end{cases}    
\label{eq:imLRELU}
\end{equation}
\vspace{-0.1in}

The corresponding non-linearity in the decoder will therefore realize the operation of an inverted modified LeakyReLU given in Equation~\ref{eq:imLRELU}. In the experiments we set $\alpha=2$.

\subsection{Residual block}
\label{invresblk}

\begin{figure}[htp!]
\vspace{-0.1in}
  \centering
  \subfigure[]{\includegraphics[width=0.22\textwidth]{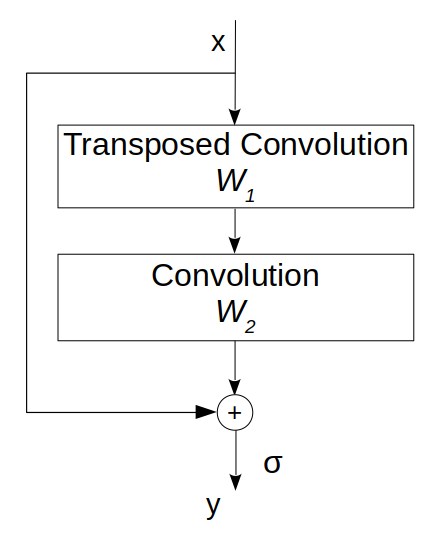} \label{fig:resnet_forward}}
  \subfigure[]{\includegraphics[width=0.22\textwidth]{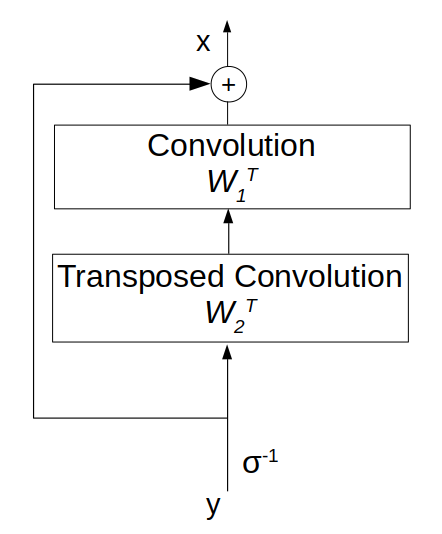}\label{fig:resnet_backward}}
  \vspace{-0.1in}
  \caption{(a) Residual block. (b) Inverted residual block.}
  \label{fig:res_block}
  \vspace{-0.05in}
\end{figure}

Consider transformation $T^E$ of an input signal performed by a residual block~\cite{resnet_2016} of an encoder $E$. We modify the residual block to remove the internal non-linearity as given in Figure~\ref{fig:res_block}a. The residual block is parametrized with weight matrices $W_1$ and $W_2$. Those are Toeplitz matrices corresponding to the convolutional and transposed convolutional layers of the residual block. Let $x$ denote block's vectorized input and $y$ denote its corresponding output. Thus transformation $T^E$ is defined as 

\vspace{-0.15in}
\begin{equation}
T^E: y = \sigma((W_2 \cdot W_1 + I)\cdot  x)
\vspace{-0.1in}
\end{equation}

An inverse operation is then defined as
\begin{equation}
(T^E)^{-1}: x = (W_2 \cdot W_1 + I)^{-1}\sigma^{-1}(y).
\label{eq:b}
\vspace{-0.15in}
\end{equation}
 
We will parametrize the counterpart residual block of the decoder with a transpose of matrix $W_2 \cdot W_1 + I$ as given in Figure~\ref{fig:res_block}b. Therefore we enforce the counterpart decoder's residual block to perform transformation:
\begin{equation}
T^{D}: x = (W_1^{\top}W_2^{\top} + I)y.
\label{eq:c}
\end{equation}
Similarly as before, at training will enforce orthonormality $(W_2 \cdot W_1 + I)^{-1} = (W_2 \cdot W_1 + I)^{\top}$ and thus $(T^E)^{-1} \equiv T^D$.

\subsection{Bias}

We consider bias as a separate layer in the network. Then, handling biases is straightforward. In particular, the layer in the encoder that perform bias addition has its counterpart layer in the decoder, where the same bias is subtracted.

\begin{figure*}[h!]
\vspace{-0.2in}
\centering
  \begin{table}[H]
  \begin{tabu} to \textwidth {X[c] X[c] X[c] X[c]}
  \large{\textbf{InvAuto}} &\large{\textbf{Auto}} &\large{\textbf{Cycle}} &\large{\textbf{VAE}}
  \end{tabu}
  \end{table}
      \vspace{-0.2in}
  \subfigure[MNIST MLP]{\includegraphics[width=0.24\textwidth]{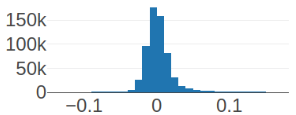} \label{1_1}}
  \subfigure[MNIST MLP]{\includegraphics[width=0.24\textwidth]{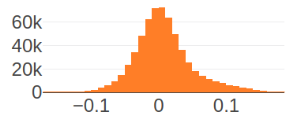} \label{1_2}}
  \subfigure[MNIST MLP]{\includegraphics[width=0.24\textwidth]{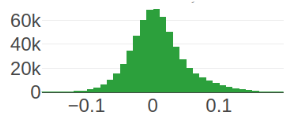} \label{1_3}}
  \subfigure[MNIST MLP]{\includegraphics[width=0.24\textwidth]{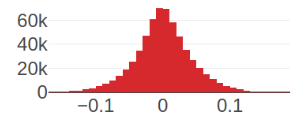} \label{1_4}}
  \subfigure[CIFAR ResNet]{\includegraphics[width=0.24\textwidth]{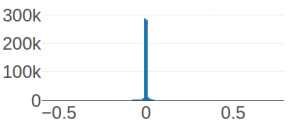} \label{4_1}}
  \subfigure[CIFAR ResNet]{\includegraphics[width=0.24\textwidth]{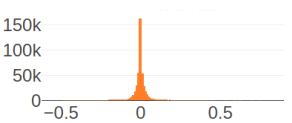} \label{4_2}}
  \subfigure[CIFAR ResNet]{\includegraphics[width=0.24\textwidth]{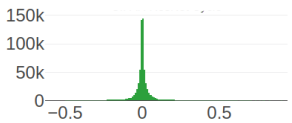} \label{4_3}}
  \subfigure[CIFAR ResNet]{\includegraphics[width=0.24\textwidth]{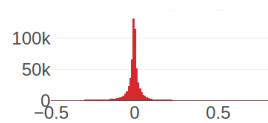} \label{4_4}}
    \vspace{-0.1in}
\caption{The histograms of cosine similarity of the rows of $E$ for InvAuto (a and e), Auto (b and f), Cycle (c and g), and VAE (d and p) on MLP and ResNet architectures and MNIST and CIFAR data sets.}
\label{fig:hist}
\vspace{-0.1in}
\end{figure*}

\subsection{Experimental validation of orthonormality} 
\label{subsec:ev}

In this section, we validate the concept of InvAuto. The goal of this section is to show that proposed shared parametrization and training enforce orthonormality and that at the same time the orthonormality property is not organically achieved by standard architectures. We compare InvAuto with previously mentioned vanilla autoencoder, autoencoder with cycle consistency, and variational autoencoder. We experimented with various data sets (MNIST and CIFAR-10) and architectures (MLP, convolutional (Conv), and ResNet). All the networks were designed to have $2$ down-sampling layers and $2$ up-sampling layers. Encoder's matrix $E$ and decoder's matrix $D$ are constructed by multiplying the weight matrices of consecutive layers of encoder and decoder, respectively.

\begin{table}[H]
\centering
\begin{tabular}{|c||c|c|c|c|}
\hline
Data set       &InvAuto & Auto & Cycle  & VAE  \\
and model     & &  &   &   \\
\hline
MNIST        & $\bm{0.001}$ & $0.008$ & $0.007$ & $0.001$  \\
MLP      & $\bm{\pm 0.118}$ & $\pm 0.210$ & $\pm 0.207$ & $\pm 0.219$  \\
\hline
MNIST           & $\bm{0.001}$ & $0.001$ & $0.001$ & $-0.001$ \\
Conv          & $\bm{\pm 0.148}$ & $\pm 0.179$ & $\pm 0.176$ & $\pm 0.190$ \\
\hline
CIFAR          & $\bm{0.001}$ & $0.002$ & $0.004$ & $0.003$ \\
Conv        & $\bm{\pm 0.145}$ & $\pm 0.176$ & $\pm 0.195$ & $\pm 0.268$ \\
\hline
CIFAR       & $\bm{0.000}$ & $0.000$ & $0.000$  & $0.001$ \\
ResNet    & $\bm{\pm 0.134}$ & $\pm 0.203$ & $\pm 0.232$  & $\pm 0.298$ \\
\hline
\end{tabular}
\vspace{-0.1in}
\caption{Mean and standard deviation of cosine similarity of rows of $E$. InvAuto achieves cosine similarity that is the closest to $0$.}
\label{tab:histtab}
\vspace{-0.15in}
\end{table}

\begin{table}[t]
\vspace{0.05in}
\centering
\begin{tabular}{|c||c|c|c|c|}
\hline
Data set       &InvAuto & Auto & Cycle  & VAE  \\
and model     & &  &   &   \\
\hline
MNIST        & $\bm{0.976}$ & $1.326$ & $1.268$ & $1.832$  \\
MLP      & $\bm{\pm 0.190}$ & $\pm 0.095$ & $\pm 0.095$ & $\pm 0.501$  \\
\hline
MNIST           & $\bm{0.905}$ & $1.699$ & $1.780$ & $1.971$ \\
Conv          & $\bm{\pm 0.321}$ & $\pm 0.732$ & $\pm 0.779$ & $\pm 0.794$ \\
\hline
CIFAR          & $\bm{0.908}$ & $3.027$ & $2.463$ & $1.176$ \\
Conv        & $\bm{\pm 0.219}$ & $\pm 0.816$ & $\pm 0.688$ & $\pm 0.356$ \\
\hline
CIFAR       & $\bm{0.868}$ & $2.890$ & $2.650$ & $1.728$ \\
ResNet    & $\bm{\pm 0.078}$ & $\pm 0.895$ & $\pm 0.937$ & $\pm 0.311$ \\
\hline
\end{tabular}
\vspace{-0.13in}
\caption{Mean and standard deviation of the $\ell_2$-norm of the rows of $E$. InvAuto achieves the $\ell_2$-norm of the rows that is the closest to the unit norm.}
\label{tab:norm}
\vspace{-0.15in}
\end{table}

We test orthonormality by reporting the histograms of the cosine similarity of each pair of rows of matrix $E$ for all methods (Figure~\ref{fig:hist}) along with their mean and standard deviation (Table~\ref{tab:histtab}) as we expect the cosine similarity to be close to $0$ for InvAuto. We then show the $\ell_2$-norm of the rows of $E$ as we expect the rows of InvAuto to have close-to-unit norm (Table~\ref{tab:norm}). InvAuto enforces the encoder, and consequently the decoder, to be orthonormal. Other methods do not explicitly demand that and thus the orthonormality of their encoders is weaker. This observation is further confirmed by Figures~\ref{fig:examplea} and~\ref{fig:exampleb} shown before in the Introduction. In the Supplement (Section~\ref{sec:A}), we provide three more figures that complement Figure~\ref{fig:exampleb} (recall that the latter reports the MSE of $DE-I$). They show the MSE of the diagonal (Figure~\ref{fig:exampleb1}) and off-diagonal of $DE-I$ (Figure~\ref{fig:exampleb2}) as well as the ratio of the MSE of the off-diagonal and diagonal of $DE$ (Figure~\ref{fig:exampleb3}) for various methods. The reconstruction loss obtained for all methods is also shown in Section~\ref{sec:A} in the Supplement (Table~\ref{tab:loss}). 

Next we describe how InvAuto is applied to the problem of domain adaptation.

\begin{figure*}[htp!]
\includegraphics[width = \textwidth]{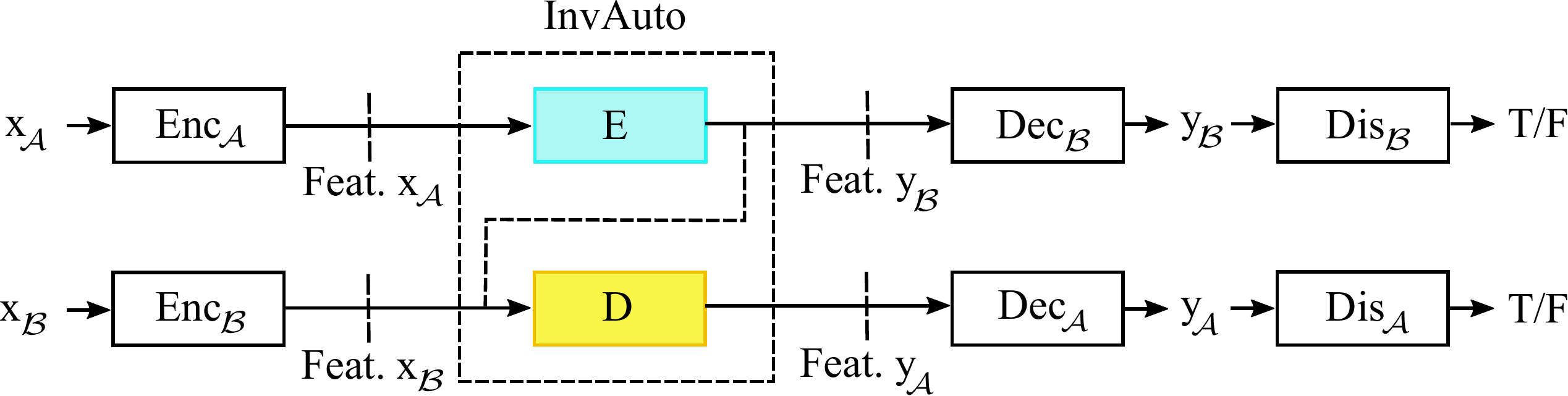}
  \centering
  \caption{The architecture of the domain translator with InvAuto $(E,D)$. $x_\mathcal{A} \in \mathcal{A}$ and $x_\mathcal{B} \in \mathcal{B}$ are the inputs of the translator. $y_{\mathcal{B}}$ is a converted image $x_\mathcal{A}$ into the $\mathcal{B}$ domain and $y_{\mathcal{A}}$ is a converted image $x_\mathcal{B}$ into the $\mathcal{A}$ domain. Invertible autoencoder $(E,D)$ is built of encoder $E$ and decoder $D$, where each of those itself is an autoencoder. $\text{Enc}_{\mathcal{A}}, \text{Enc}_{\mathcal{B}}$ are feature extractors, and $\text{Dec}_{\mathcal{A}}, \text{Dec}_{\mathcal{B}}$ are the final layers of the generators $\text{Gen}_{\mathcal{B}}$, i.e. ($\text{Enc}_{\mathcal{A}},E,\text{Dec}_{\mathcal{B}}$), and $\text{Gen}_{\mathcal{A}}$, i.e. ($\text{Enc}_{\mathcal{B}},D,\text{Dec}_{\mathcal{A}}$), respectively. Discriminators $\text{Dis}_{\mathcal{A}}$ and $\text{Dis}_{\mathcal{B}}$ discriminate whether their input comes from the generator (\textbf{T}rue) or original data set (\textbf{F}alse).}
  \label{fig:InvAuto}
  \vspace{-0.05in}
\end{figure*}

\section{Invertible autoencoder for domain adaptation}
\label{sec:iada}

For the purpose of performing domain adaptation we construct the dedicated architecture that is similar to CycleGAN, but we use InvAuto at the feature level of the generators. This InvAuto contains encoder $E$ and decoder $D$ that themselves have the form of autoencoders. Each of these internal autoencoders is used to do the conversion between the features corresponding to two different domains. And thus, the encoder $E$ performs the conversion from the features corresponding to domain $\mathcal{A}$ into the features corresponding to domain $\mathcal{B}$. The decoder $D$, on the other hand, performs the conversion from the features corresponding to domain $\mathcal{B}$ into the features corresponding to domain $\mathcal{A}$. Since $E$ and $D$ form InvAuto, $E$ realizes an inversion of $D$ (and vice versa) and shares parameters with $D$. This introduces strong correlations between two generators and reduces the number of trainable parameters, which distinguishes our approach from CycleGAN. The proposed architecture is illustrated in Figure~\ref{fig:InvAuto}. The details of the architecture and training are provided in Section~\ref{sec:C} in the Supplement.

Next we describe the cost function that we use to train our deep model. The first component of the cost function is the adversarial loss~\cite{gan_ian}, i.e. 
\begin{equation}
\vspace{-0.1in}
\begin{split}
L_{\text{adv}}(\text{Gen}_{\mathcal{A}},\text{Dis}_{\mathcal{A}}) =& \mathbb{E}_{x_{\mathcal{A}}\sim p_{d}({\mathcal{A}})}[\log \text{Dis}_{\mathcal{A}}(x_{\mathcal{A}})]+\\
					& \mathbb{E}_{x_{\mathcal{B}}\sim p_{d}({\mathcal{B}})}[\log (1 - \text{Dis}_{\mathcal{A}}(\text{Gen}_{\mathcal{A}}(x_{\mathcal{B}})))]\\
L_{\text{adv}}(\text{Gen}_{\mathcal{B}},\text{Dis}_{\mathcal{B}}) =& \mathbb{E}_{x_{\mathcal{B}}\sim p_{d}({\mathcal{B}})}[\log \text{Dis}_{\mathcal{B}}(x_{\mathcal{B}})]+\\
					& \mathbb{E}_{x_{\mathcal{A}}\sim p_{d}({\mathcal{A}})}[\log (1 - \text{Dis}_{\mathcal{B}}(\text{Gen}_{\mathcal{B}}(x_{\mathcal{A}})))],
\end{split}
\end{equation}
where $p_{d}({\mathcal{A}})$ and $p_{d}({\mathcal{B}})$ denote the distribution of data from $\mathcal{A}$ and $\mathcal{B}$, respectively.

The second component of the loss function is the cycle consistency loss defined as

\vspace{-0.2in}
\begin{equation}
\begin{split}
L_{cc}(\text{Gen}_{\mathcal{A}}, \text{Gen}_{\mathcal{B}}) &= \mathbb{E}_{x_{\mathcal{A}}\sim p_{d}(\mathcal{A})}
						[\lVert {x_A - \text{Gen}_{\mathcal{A}}(\text{Gen}_{\mathcal{B}}(x_A))}\rVert_1 ]\\
					 &+ \mathbb{E}_{x_{\mathcal{B}}\sim p_{d}({\mathcal{B}})}
						[\lVert {x_B - \text{Gen}_{\mathcal{B}}(\text{Gen}_{\mathcal{A}}(x_B))}\rVert_1 ].                    
\end{split}
\end{equation}
\vspace{-0.2in}

The objective function that we minimize therefore becomes
\begin{equation}
\label{cycle_loss}
\begin{split}
L(\text{Gen}_{\mathcal{A}}, \text{Gen}_{\mathcal{B}}, \text{Dis}_{\mathcal{A}}, \text{Dis}_{\mathcal{B}}) &=\lambda L_{cc}(\text{Gen}_{\mathcal{A}}, \text{Gen}_{\mathcal{B}})\\
							  &+ L_{\text{adv}}(\text{Gen}_{\mathcal{A}},\text{Dis}_{\mathcal{A}})\\ 
                              &+ L_{\text{adv}}(\text{Gen}_{\mathcal{B}},\text{Dis}_{\mathcal{B}}),
\end{split}
\end{equation}
where $\lambda$ controls the balance between the adversarial loss and cycle consistency loss. The cycle consistency loss enforces the orthonormality property of InvAuto.

\section{Experiments} 
\label{Experiments}

We next demonstrate the experiments on domain adaptation problems. We compare our model against UNIT\cite{unit} and CycleGAN\cite{CycleGAN2017}. We used publicly available implementations of both methods available from \url{https://github.com/mingyuliutw/UNIT/} and \url{https://github.com/junyanz/pytorch-CycleGAN-and-pix2pix/}.The details of our architecture and the training process are summarized in Section~\ref{sec:C} in the Supplement. 

\subsection{Experiments with benchmark data sets}

We considered the following domain adaptation tasks:
\begin{itemize}
\item[(i)] Day-to-night and night-to-day image conversion: we used unpaired road pictures recorded during the day and at night obtained from KAIST data set~\cite{kaist}. 

\item[(ii)] Day-to-thermal and thermal-to-day image conversion: we used road pictures recorded during the day with a regular camera and a thermal camera obtained from KAIST data set\cite{kaist}. 

\item[(iii)] Maps-to-satellite and satellite-to-maps: we used satellite images and maps obtained from Google Maps~\cite{pix2pix}. 
\end{itemize}
\vspace{-0.05in}

The data sets for the last two tasks, i.e. (ii) and (iii), are originally paired, however we randomly permuted them and train the model in an unsupervised fashion. The training and testing images were furthermore resized to $128 \times 128$ resolution. 

The visual results of image conversion are presented in Figures~\ref{fig:conv1}-\ref{fig:conv6} (Section~\ref{sec:B} in the Supplement contains the same figures in higher resolution). We see that InvAuto visually performs comparably to other state-of-the-art methods.

\begin{figure}[H]
\vspace{-0.05in}
\begin{tabu} to 0.48\textwidth {X[c] X[c] X[c] X[c]}
\textbf{Original} & \textbf{CycleGAN} & \textbf{UNIT} & \textbf{InvAuto}
\end{tabu}
\includegraphics[width=0.48\textwidth]{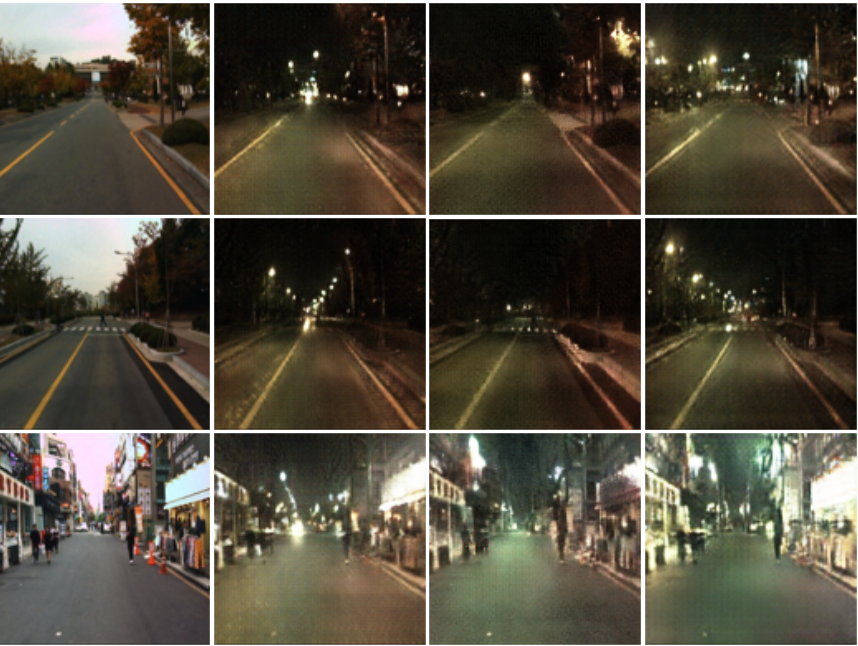}
\vspace{-0.3in}
\caption{Day-to-night image conversion. Zoomed image is shown in Figure~\ref{fig:conv1supp} in Section~\ref{sec:B} of the Supplement.}
\label{fig:conv1}
\end{figure}

\begin{figure}[H]
\vspace{-0.15in}
\begin{tabu} to 0.48\textwidth {X[c] X[c] X[c] X[c]}
\textbf{Original} & \textbf{CycleGAN} & \textbf{UNIT} & \textbf{InvAuto}
\end{tabu}
\includegraphics[width=0.48\textwidth]{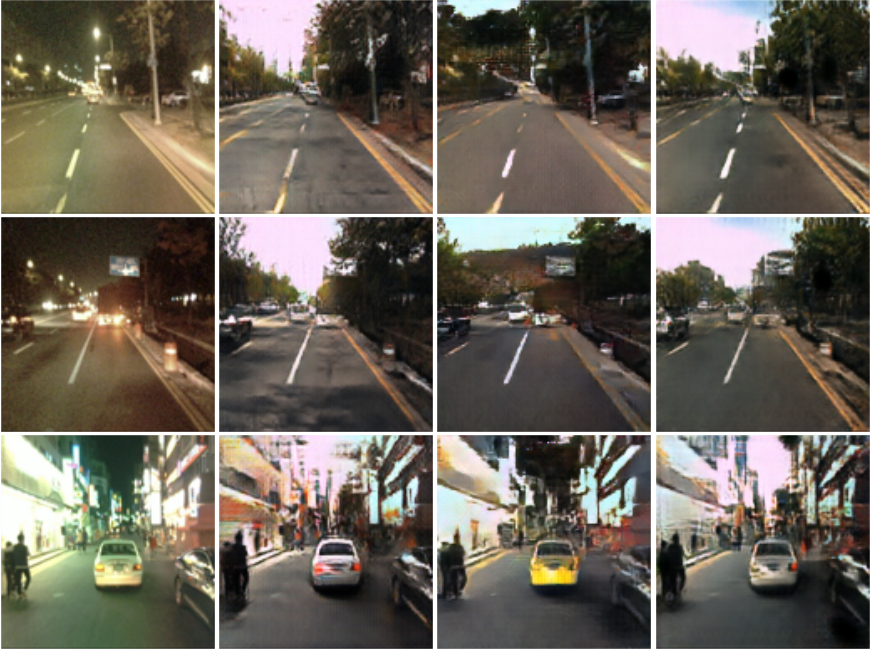}
\vspace{-0.3in}
\caption{Night-to-day image conversion. Zoomed image is shown in Figure~\ref{fig:conv2supp} in Section~\ref{sec:B} of the Supplement.}
\label{fig:conv2}
\end{figure}

\begin{figure}[H]
\begin{tabu} to 0.47\textwidth {X[c] X[c] X[c] X[c] X[c]}
\textbf{Original} & \textbf{CycleGAN} & \textbf{UNIT} & \textbf{InvAuto} & \textbf{Reference}
\end{tabu}
\includegraphics[width=0.48\textwidth, height=1.85in]{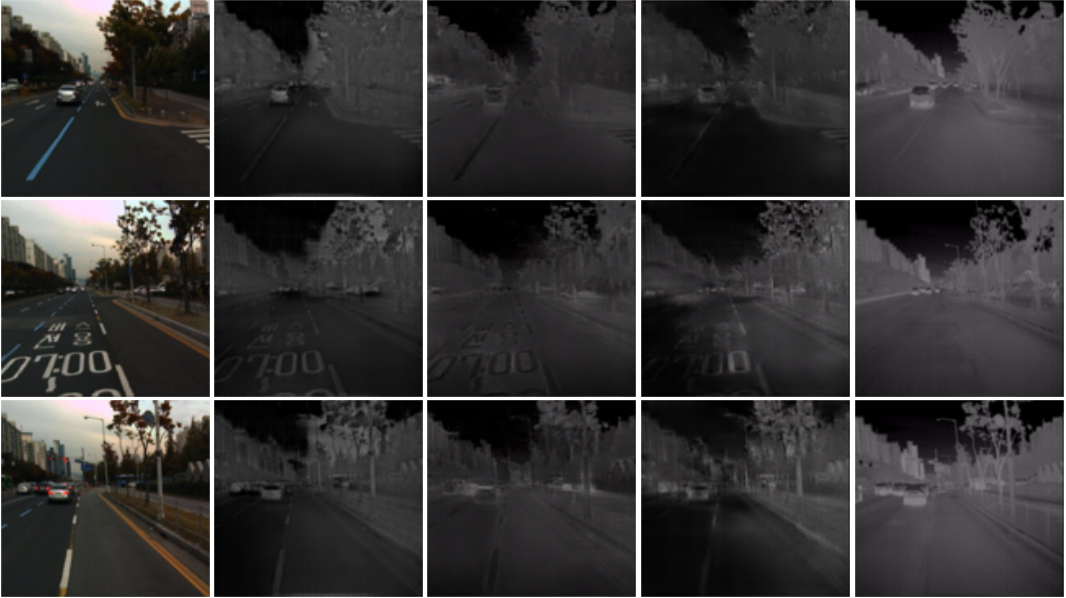}
\vspace{-0.25in}
\caption{Day-to-thermal image conversion. Zoomed image is shown in Figure~\ref{fig:conv3supp} in Section~\ref{sec:B} of the Supplement.}
\label{fig:conv3}
\end{figure}

\begin{figure}[H]
\vspace{-0.25in}
\begin{tabu} to 0.47\textwidth {X[c] X[c] X[c] X[c] X[c]}
\textbf{Original} & \textbf{CycleGAN} & \textbf{UNIT} & \textbf{InvAuto} & \textbf{Reference}
\end{tabu}
\includegraphics[width=0.48\textwidth, height=1.85in]{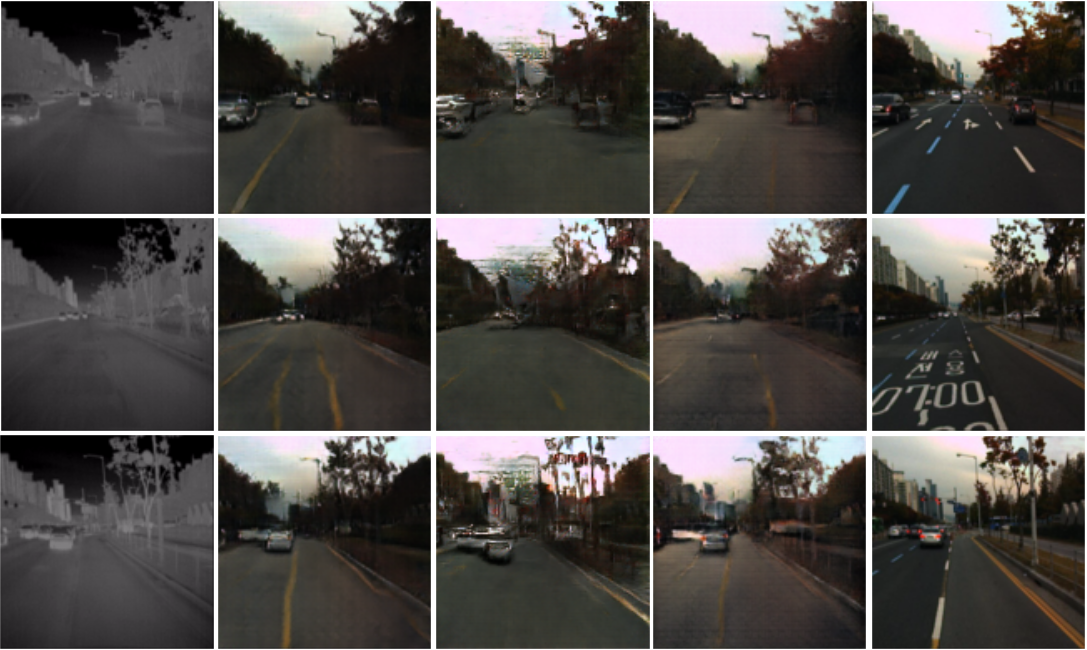}
\label{fig:conv4}
\vspace{-0.25in}
\caption{Thermal-to-day image conversion. Zoomed image is shown in Figure~\ref{fig:conv4supp} in Section~\ref{sec:B} of the Supplement.}
\vspace{-0.2in}
\end{figure}

To evaluate the performance of the methods numerically we use the following approach:
\vspace{-0.05in}
\begin{itemize}
\setlength\itemsep{0em}
\item For the tasks (ii) and (iii), we directly calculated the $\ell_1$ loss between the converted images and the ground truth. 

\item For the task (i), we trained two autoencoders $\Omega_{\mathcal{A}}$ and $\Omega_{\mathcal{B}}$ on both domains, i.e. we trained each of them to reconstruct well the images from its own domain and reconstruct badly the images from the other domain. Then we use these two autoencoders
to evaluate the quality of the converted images, where high $\ell_1$ reconstruction loss of the autoencoder for the images converted to resemble those from its corresponding domain implies low-quality image translation. 
\end{itemize}
\vspace{-0.05in}

Table~\ref{tab:nc} contains the results of the numerical evaluation and shows that the performance of InvAuto is similar to the state-of-the-art techniques that we compare InvAuto with and is furthermore contained within the performance range established by the CycleGAN (best performer) and UNIT (consistently slightly worst from CycleGAN).

\begin{figure}[h]
\begin{tabu} to 0.47\textwidth {X[c] X[c] X[c] X[c] X[c]}
\textbf{Original} & \textbf{CycleGAN} & \textbf{UNIT} & \textbf{InvAuto} & \textbf{Reference}
\end{tabu}
\includegraphics[width=0.48\textwidth, height=1.85in]{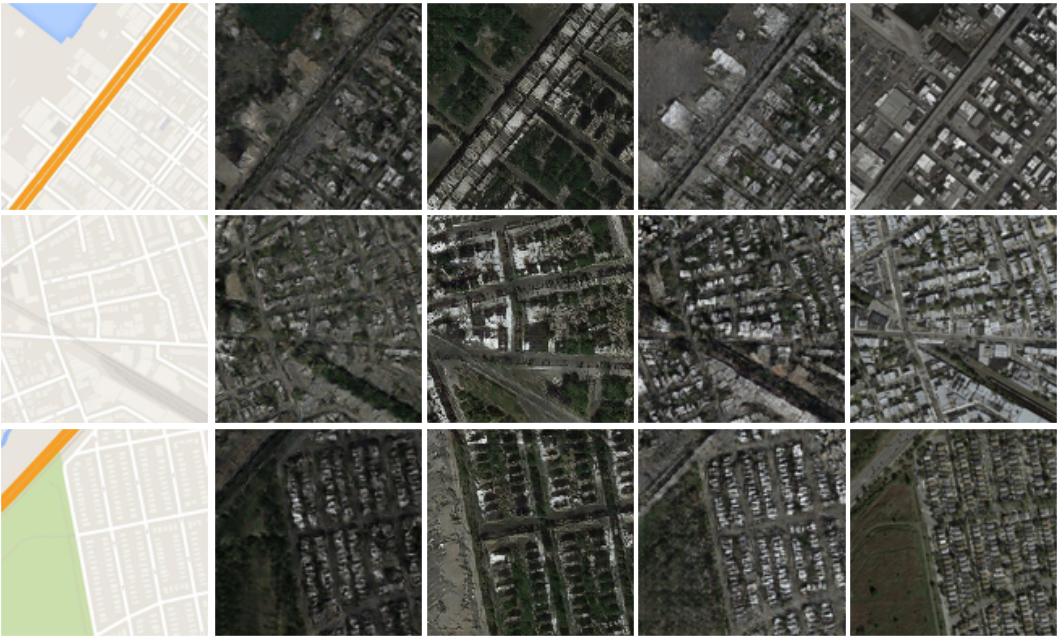}
\vspace{-0.25in}
\caption{Maps-to-satellite image conversion. Zoomed image is shown in Figure~\ref{fig:conv5supp} in Section~\ref{sec:B} of the Supplement.}
\label{fig:conv5}
\end{figure}

\begin{figure}[h]
\begin{tabu} to 0.47\textwidth {X[c] X[c] X[c] X[c] X[c]}
\textbf{Original} & \textbf{CycleGAN} & \textbf{UNIT} & \textbf{InvAuto} & \textbf{Reference}
\end{tabu}
\includegraphics[width=0.48\textwidth, height=1.85in]{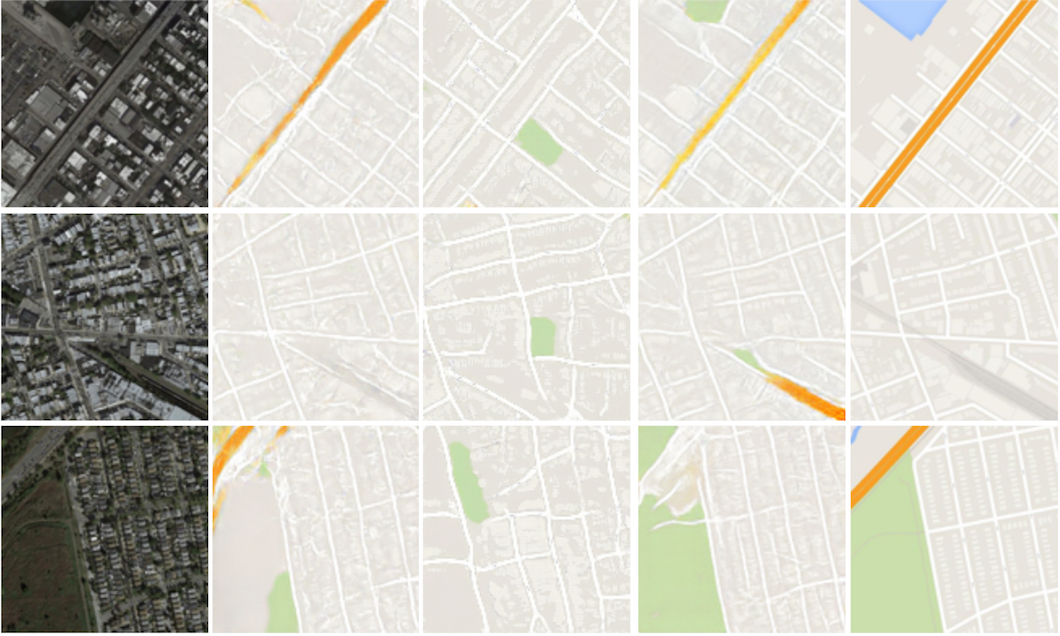}
\vspace{-0.25in}
\caption{Satellite-to-maps image conversion. Zoomed image is shown in Figure~\ref{fig:conv6supp} in Section~\ref{sec:B} of the Supplement.}
\label{fig:conv6}
\end{figure}
\vspace{-0.05in}

\begin{table}[H]
\vspace{-0.15in}
\centering
\begin{tabular}{c|c|c|c|}
\cline{2-4}
            & \multicolumn{3}{|c|}{\multirow{1}{*}{Methods}}   \\
\hline
\multicolumn{1}{|c||}{\multirow{1}{*}{Tasks}}            & CycleGAN & UNIT & InvAuto  \\
\hline
\multicolumn{1}{|c||}{\multirow{1}{*}{Night-to-day}}             & $0.033$ & $0.227$ & $0.062$  \\
\hline
\multicolumn{1}{|c||}{\multirow{1}{*}{Day-to-nigth}}             & $0.041$ & $0.114$ & $0.067$  \\
\hline
\multicolumn{1}{|c||}{\multirow{1}{*}{Thermal-to-day}}           & $0.287$ & $0.339$ & $0.299$  \\
\hline
\multicolumn{1}{|c||}{\multirow{1}{*}{Day-to-thermal}}           & $0.179$ & $0.194$ & $0.205$  \\
\hline
\multicolumn{1}{|c||}{\multirow{1}{*}{Maps-to-satellite}}   & $0.261$ & $0.331$ & $0.272$  \\
\hline
\multicolumn{1}{|c||}{\multirow{1}{*}{Satellite-to-maps}}   & $0.069$ & $0.104$ & $0.080$  \\
\hline
\end{tabular}
\vspace{-0.05in}
\caption{Numerical evaluation of CycleGAN, UNIT, and InvAuto.}
\label{tab:nc}
\end{table}

\subsection{Experiments with autonomous driving system} 

\begin{figure}[t]
\begin{tabu} to 0.48\textwidth {X[c] X[c] X[c]}
\textbf{Original} & \textbf{CycleGAN} & \textbf{InvAuto} 
\end{tabu}
\includegraphics[width=0.48 \textwidth]{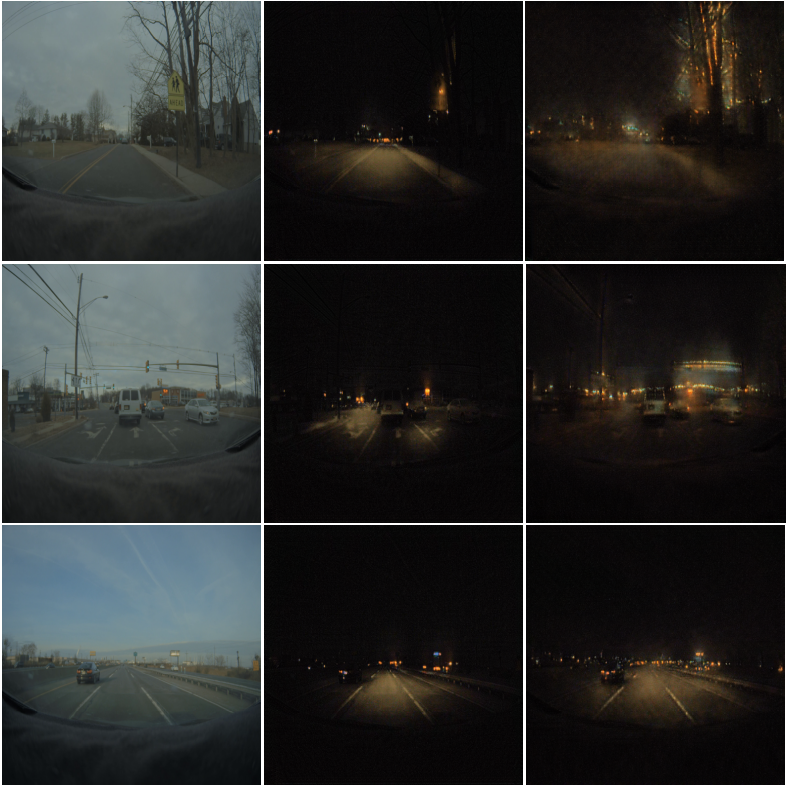}
\vspace{-0.25in}
  \caption{Experimental results with autonomous driving system: day-to-night conversion. Zoomed image is shown in Figure~\ref{fig:car1conv1} in Section~\ref{sec:B} of the Supplement.}
  \centering
  \label{fig:car1conv}
\end{figure}

\begin{figure}[h!]
\begin{tabu} to 0.48\textwidth {X[c] X[c] X[c]}
\textbf{Original} & \textbf{CycleGAN} & \textbf{InvAuto} 
\end{tabu}
\includegraphics[width=0.48\textwidth]{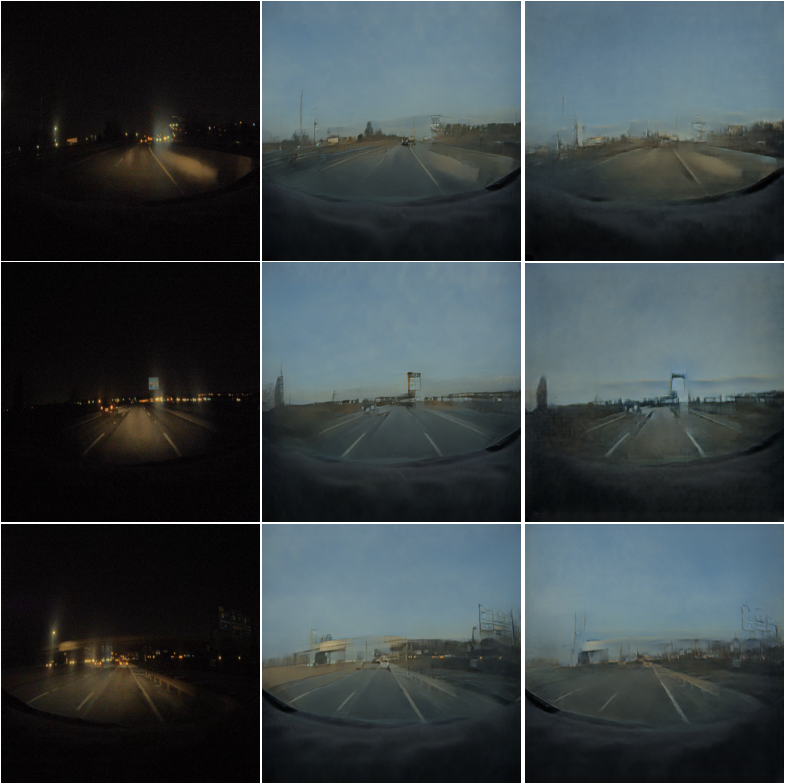}
\vspace{-0.25in}
  \caption{Experimental results with autonomous driving system: night-to-day conversion. Zoomed image is shown in Figure~\ref{fig:car2conv2} in Section~\ref{sec:B} of the Supplement.}
  \centering
  \label{fig:car2conv}
\end{figure}

To test the quality of the image-to-image translations obtained by InvAuto, we use the NVIDIA evaluation system for autonomous driving described in details in~\cite{DBLP:journals/corr/BojarskiTDFFGJM16}. The system evaluates the performance of an already trained NVIDIA neural-network-based end-to-end learning platform for autonomous driving (PilotNet) on a test video using a simulator for autonomous driving. The system uses the following performance metrics for evaluation: autonomy, position precision, and comfort. We do not describe these metrics as they are described well in the mentioned paper. We only emphasize that these metrics are expressed as a percentage, where $100\%$ corresponds to the best performance. We collected the high-resolution videos of the same road during the day and night from the camera inside the car. Each video had $\sim 45$K frames. The pictures were resized to $512 \times 512$ resolution for the conversion and then resized back to the original size of $1920\times 1208$. We used our domain translator as well as CycleGAN to convert the collected day video to a night video (Figure~\ref{fig:car1conv}) and also the collected night video to a day video (Figure~\ref{fig:car2conv}). To evaluate our model, we used aforementioned NVIDIA evaluation system, where the converted videos where used as testing sets for this system. We report results in Table~\ref{tab:car1}. 

\begin{table}[H]
\vspace{-0.15in}
\centering
\begin{tabular}{c|c|c|c|}
\hline
\multicolumn{1}{|c||}{\multirow{1}{*}{Video type}}            & Autonomy & Position & Comfort  \\
\multicolumn{1}{|c||}{\multirow{1}{*}{}}            &  & precision &   \\
\hline
\hline
\multicolumn{1}{|c||}{\multirow{1}{*}{Original day}}          & $99.6\%$  &$73.3\%$  & $89.7\%$\\
\hline
\multicolumn{1}{|c||}{\multirow{1}{*}{Original night}}        & $98.6\%$  &$63.1\%$  & $86.3\%$\\
\hline
\multicolumn{1}{|c||}{\multirow{1}{*}{Day-to-night}}          & $99.0\%$  &$69.6\%$  & $83.2\%$\\
\multicolumn{1}{|c||}{\multirow{1}{*}{\textbf{InvAuto}}}          &   & &\\
\hline
\multicolumn{1}{|c||}{\multirow{1}{*}{Night-to-day}}          & $99.3\%$ &$68.0\%$  & $84.7\%$\\
\multicolumn{1}{|c||}{\multirow{1}{*}{\textbf{InvAuto}}}         &   & &\\
\hline
\multicolumn{1}{|c||}{\multirow{1}{*}{Day-to-night}}          & $99.0\%$  &$68.4\%$  & $84.7\%$\\
\multicolumn{1}{|c||}{\multirow{1}{*}{\textbf{CycleGAN}}}          &   & &\\
\hline
\multicolumn{1}{|c||}{\multirow{1}{*}{Night-to-day}}          & $98.8\%$ &$64.0\%$  & $87.3\%$\\
\multicolumn{1}{|c||}{\multirow{1}{*}{\textbf{CycleGAN}}}          &   & &\\
\hline
\end{tabular}
\vspace{-0.05in}
\caption{Experimental results with autonomous driving system: autonomy, position precision, and comfort.}
\label{tab:car1}
\vspace{-0.2in}
\end{table}

The PilotNet model used for testing was trained mostly on day videos, thus it is expected to perform worse on night videos. Therefore the performance for original night video is worse than for the same video converted to a day video in terms of autonomy and position precision. The comfort deteriorates due to the inconsistency of consecutive frames in the converted video, i.e. the videos are converted frame-by-frame and we do not apply any post-processing to ensure smooth transition between frames. The results for InvAuto and CycleGAN are comparable.

\section{Conclusion}
\label{sec:c}
 
We proposed a novel architecture that we call invertible autoencoder, which, as opposed to the common deep learning architectures, allows the layers of the model performing opposite operations (like encoder and decoder) to share weights. This is achieved by enforcing orthonormal mappings in the layers of the model. We demonstrate the applicability of the proposed architecture to the problem of domain adaptation and evaluate it on benchmark data sets and autonomous driving task. The performance of the proposed approach matches state-of-the-art methods and requires less trainable parameters.

\bibliography{main}
\bibliographystyle{icml2017}

\clearpage
\normalsize

\appendix

\toptitlebar 
{\Large \bf  \centering{Invertible Autoencoder for domain adaptation\\ (Supplementary material)} \par}
\bottomtitlebar

\vspace{-0.3in}

\section{Additional plots and tables for Section~\ref{subsec:ev}}
\label{sec:A}

\begin{figure}[htp!]
\vspace{-0.2in}
\centering
\includegraphics[width=0.5 \textwidth]{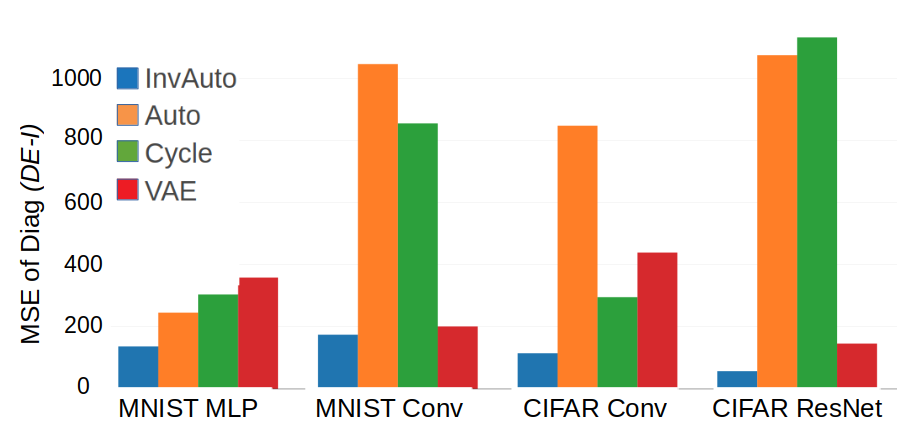}
\vspace{-0.3in}
\caption{Comparison of the MSE of the diagonal of $DE - I$ for InvAuto, Auto, Cycle, and VAE on MLP, convolutional (Conv), and ResNet architectures and MNIST and CIFAR data sets.}
\label{fig:exampleb1}
\end{figure}

\begin{figure}[htp!]
\centering
\includegraphics[width=0.5 \textwidth]{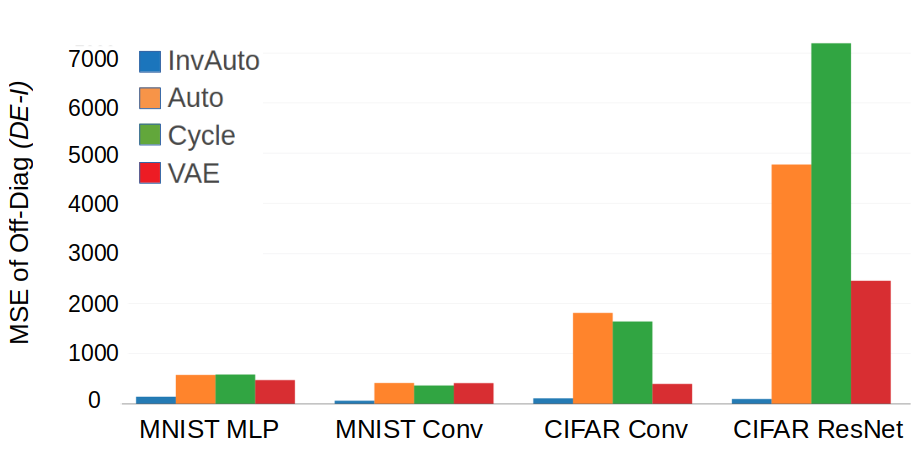}
\vspace{-0.3in}
\caption{Comparison of the MSE of the off-diagonal of $DE - I$ for InvAuto, Auto, Cycle, and VAE on MLP, convolutional (Conv), and ResNet architectures and MNIST and CIFAR data set.}
\label{fig:exampleb2}
\end{figure}

\begin{figure}[htp!]
\centering
\includegraphics[width=0.5 \textwidth]{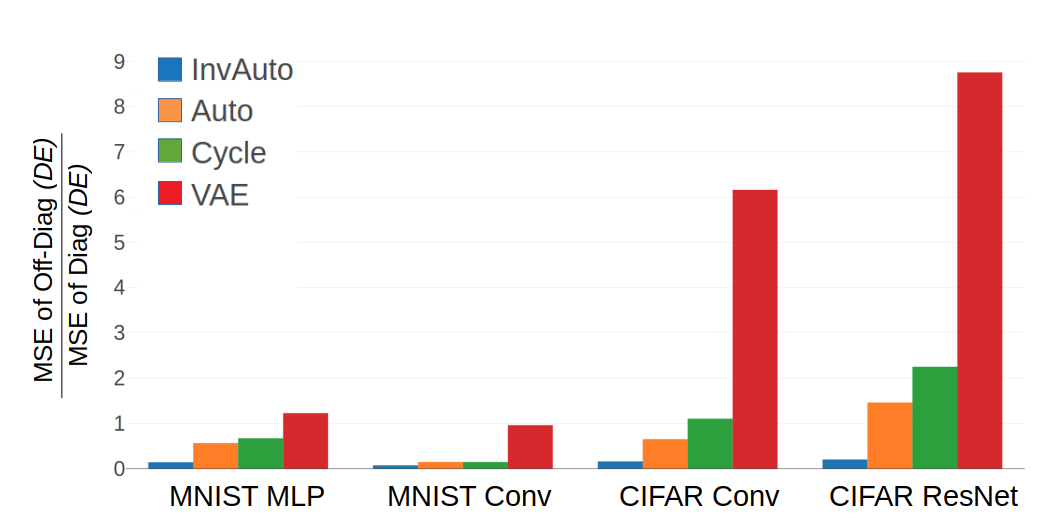}
\vspace{-0.3in}
\caption{Comparison of the ratio of MSE of the off-diagonal and diagonal of $DE$ for InvAuto, Auto, Cycle, and VAE on MLP, convolutional (Conv), and ResNet architectures and MNIST and CIFAR data sets.}
\label{fig:exampleb3}
\end{figure}

\begin{table}[H]
\vspace{0.12in}
\centering
\begin{tabular}{|c||c|c|c|c|}
\hline
Data set       &InvAuto & Auto & Cycle  & VAE  \\
and model     & &  &   &   \\
\hline
MNIST      & $0.189$ & $0.100$ & $0.112$ & $1.245$  \\
MLP & & & & \\
\hline
MNIST        & $0.168$ & $0.051$ & $0.057$ & $1.412$ \\
Conv & & & & \\
\hline
CIFAR        & $0.236$ & $0.126$ & $0.195$ & $1.457$ \\
Conv & & & & \\
\hline
CIFAR      & $0.032$ & $0.127$ & $0.217$ & $0.964$ \\
ResNet & & & & \\
\hline
\end{tabular}
\caption{Test reconstruction loss (MSE) for InvAuto, Auto, Cycle, and VAE on MLP, convolutional (Conv), and ResNet architectures and MNIST and CIFAR data sets. VAE has significantly higher reconstruction loss by construction.}
\label{tab:loss}
\end{table}

\begin{figure*}[htp!]
\centering
  \begin{table}[H]
  \begin{tabu} to \textwidth {X[c] X[c] X[c] X[c]}
  \large{\textbf{InvAuto}} &\large{\textbf{Auto}} &\large{\textbf{Cycle}} &\large{\textbf{VAE}}
  \end{tabu}
  \end{table}
  \vspace{-0.2in}
  \subfigure[MNIST MLP]{\includegraphics[width=0.24\textwidth]{{1-1}.jpg} \label{1_1}}
  \subfigure[MNIST MLP]{\includegraphics[width=0.24\textwidth]{{1-2}.jpg} \label{1_2}}
  \subfigure[MNIST MLP]{\includegraphics[width=0.24\textwidth]{{1-3}.jpg} \label{1_3}}
  \subfigure[MNIST MLP]{\includegraphics[width=0.24\textwidth]{{1-4}.jpg} \label{1_4}}
  \subfigure[MNIST Conv]{\includegraphics[width=0.24\textwidth]{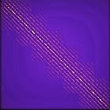} \label{2_1}}
  \subfigure[MNIST Conv]{\includegraphics[width=0.24\textwidth]{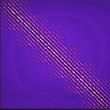} \label{2_2}}
  \subfigure[MNIST Conv]{\includegraphics[width=0.24\textwidth]{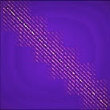} \label{2_3}}
  \subfigure[MNIST Conv]{\includegraphics[width=0.24\textwidth]{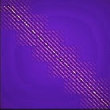} \label{2_4}}
  \subfigure[CIFAR Conv]{\includegraphics[width=0.24\textwidth]{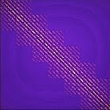} \label{3_1}}
  \subfigure[CIFAR Conv]{\includegraphics[width=0.24\textwidth]{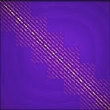} \label{3_2}}
  \subfigure[CIFAR Conv]{\includegraphics[width=0.24\textwidth]{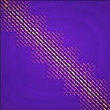} \label{3_3}}
  \subfigure[CIFAR Conv]{\includegraphics[width=0.24\textwidth]{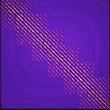} \label{3_4}}
  \subfigure[CIFAR ResNet]{\includegraphics[width=0.24\textwidth]{{4-1}.jpg} \label{4_1}}
  \subfigure[CIFAR ResNet]{\includegraphics[width=0.24\textwidth]{{4-2}.jpg} \label{4_2}}
  \subfigure[CIFAR ResNet]{\includegraphics[width=0.24\textwidth]{{4-3}.jpg} \label{4_3}}
  \subfigure[CIFAR ResNet]{\includegraphics[width=0.24\textwidth]{{4-4}.jpg} \label{4_4}}
\caption{Heatmap of the values of matrix $DE$ for InvAuto, (a,e,i,m) Auto (b,f,j,n), Cycle (c,g,k,o), and VAE (d,h,l,p) on MLP, convolutional (Conv), and ResNet architectures and MNIST and CIFAR data sets. Matrices $E$ and $D$ are constructed by multiplying the weight matrices of consecutive layers of encoder and decoder, respectively. In case of InvAuto, $DE$ is the closest to the identity matrix.}
\label{fig:exampleasup}
\end{figure*}

\newpage
\onecolumn

\section{Additional experimental results for Section~\ref{Experiments}}
\label{sec:B}

\begin{table}[H]
\centering
\begin{tabu} to \textwidth {X[c] X[c] X[c] X[c]}
\large{\textbf{Original}} & \large{\textbf{CycleGAN}} & \large{\textbf{UNIT}} & \large{\textbf{InvAuto}}
\end{tabu}
\vspace{-0.3in}
\begin{figure}[H]
\centering
\includegraphics[width=\textwidth]{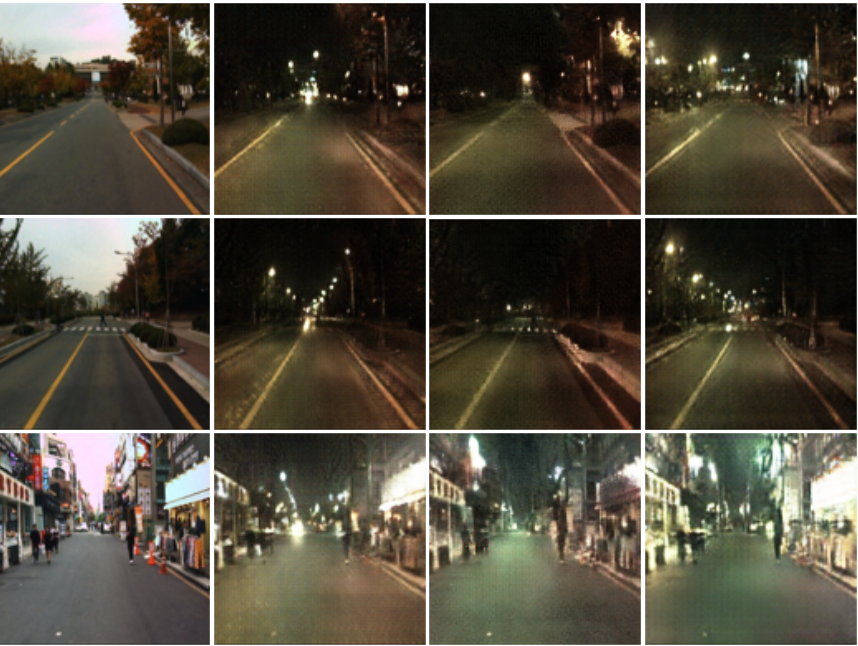}
\vspace{-0.2in}
\caption{Day-to-night image conversion.}
\label{fig:conv1supp}
\end{figure}
\end{table}

\newpage
\begin{table}[H]
\centering
\vspace{0.15in}
\begin{tabu} to \textwidth {X[c] X[c] X[c] X[c]}
\large{\textbf{Original}} & \large{\textbf{CycleGAN}} & \large{\textbf{UNIT}} & \large{\textbf{InvAuto}}
\end{tabu}
\vspace{-0.3in}
\begin{figure}[H]
\centering
\includegraphics[width=\textwidth]{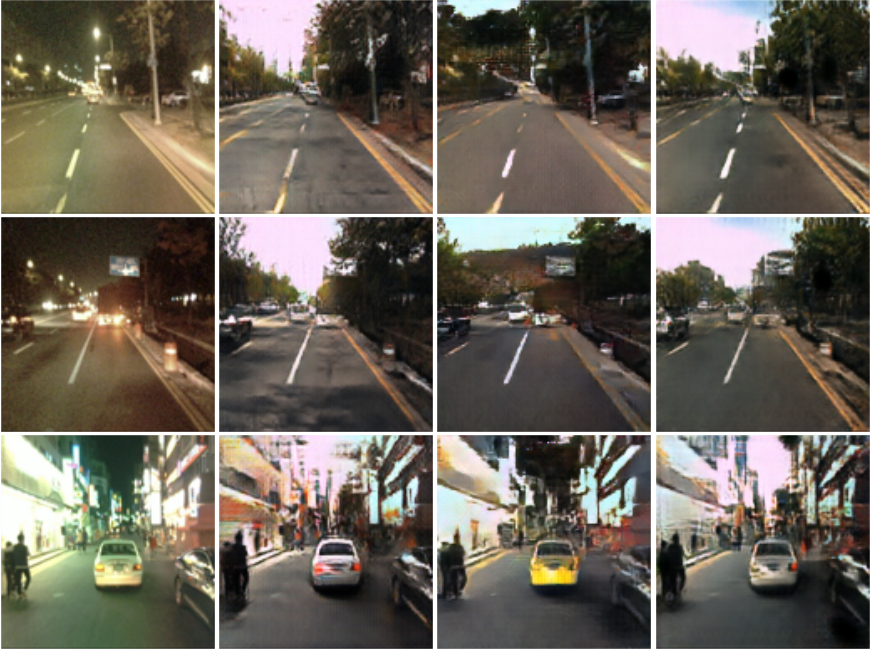}
\vspace{-0.2in}
\caption{Night-to-day image conversion.}
\label{fig:conv2supp}
\end{figure}
\end{table}

\newpage
\begin{table}[H]
\centering
\vspace{0.15in}
\begin{tabu} to \textwidth {X[c] X[c] X[c] X[c] X[c]}
\large{\textbf{Original}} & \large{\textbf{CycleGAN}} & \large{\textbf{UNIT}} & \large{\textbf{InvAuto}} & \large{\textbf{Reference}}
\end{tabu}
\vspace{-0.3in}
\begin{figure}[H]
\centering
\includegraphics[width=\textwidth]{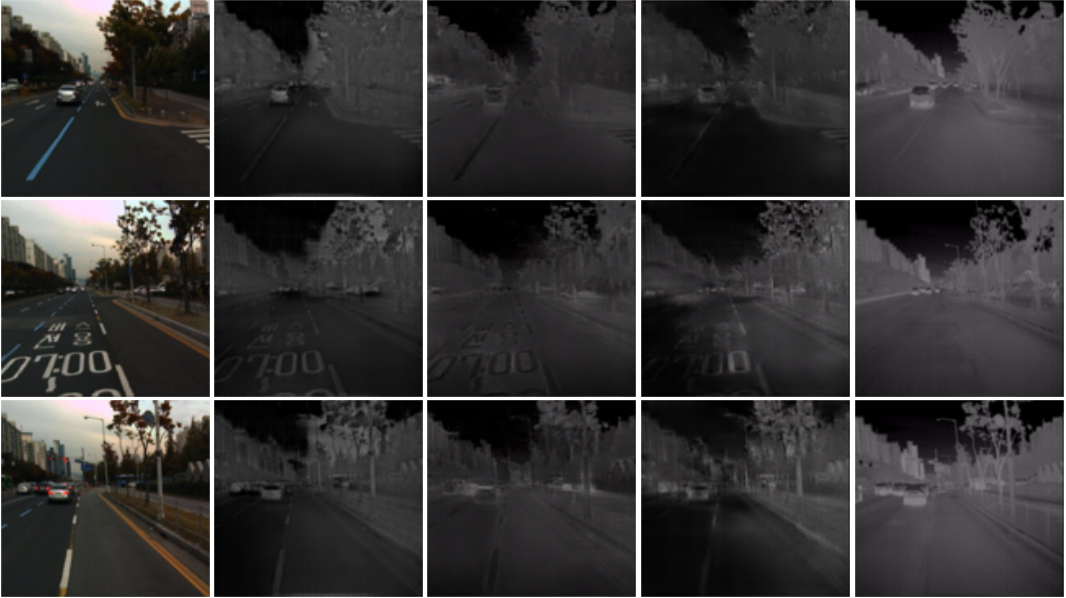}
\vspace{-0.2in}
\caption{Day-to-thermal image conversion.}
\label{fig:conv3supp}
\end{figure}
\end{table}

\newpage
\begin{table}[H]
\centering
\vspace{0.15in}
\begin{tabu} to \textwidth {X[c] X[c] X[c] X[c] X[c]}
\large{\textbf{Original}} & \large{\textbf{CycleGAN}} & \large{\textbf{UNIT}} & \large{\textbf{InvAuto}} & \large{\textbf{Reference}}
\end{tabu}
\vspace{-0.3in}
\begin{figure}[H]
\centering
\includegraphics[width=\textwidth]{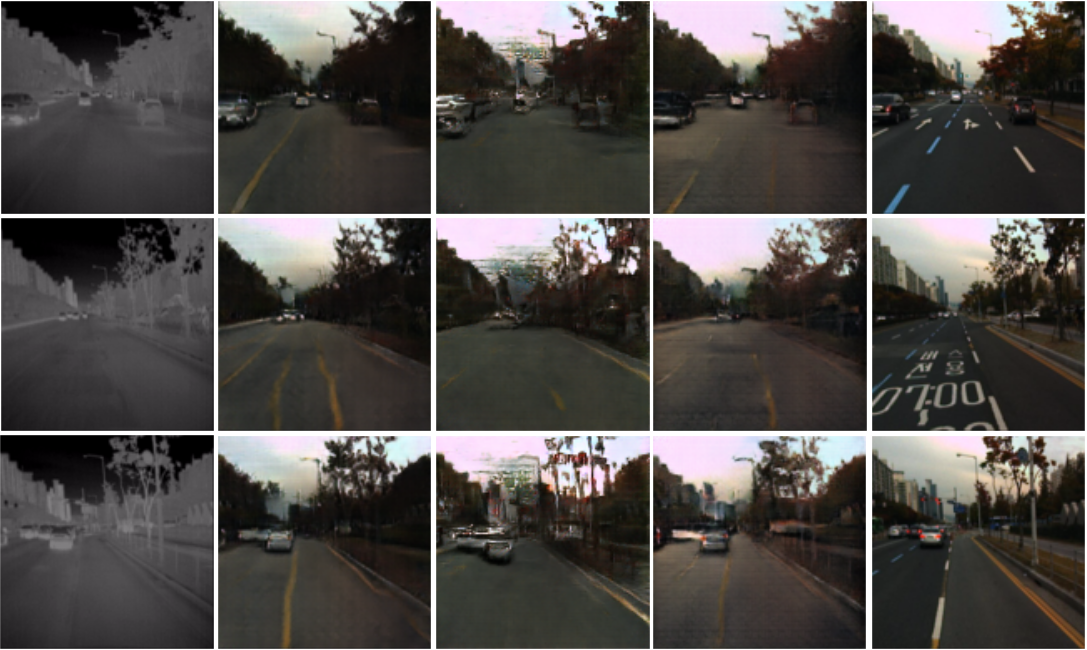}
\vspace{-0.2in}
\caption{Thermal-to-day image conversion.}
\label{fig:conv4supp}
\end{figure}
\end{table}

\newpage
\begin{table}[H]
\centering
\vspace{0.15in}
\begin{tabu} to \textwidth {X[c] X[c] X[c] X[c] X[c]}
\large{\textbf{Original}} & \large{\textbf{CycleGAN}} & \large{\textbf{UNIT}} & \large{\textbf{InvAuto}} & \large{\textbf{Reference}}
\end{tabu}
\vspace{-0.3in}
\begin{figure}[H]
\centering
\includegraphics[width=\textwidth]{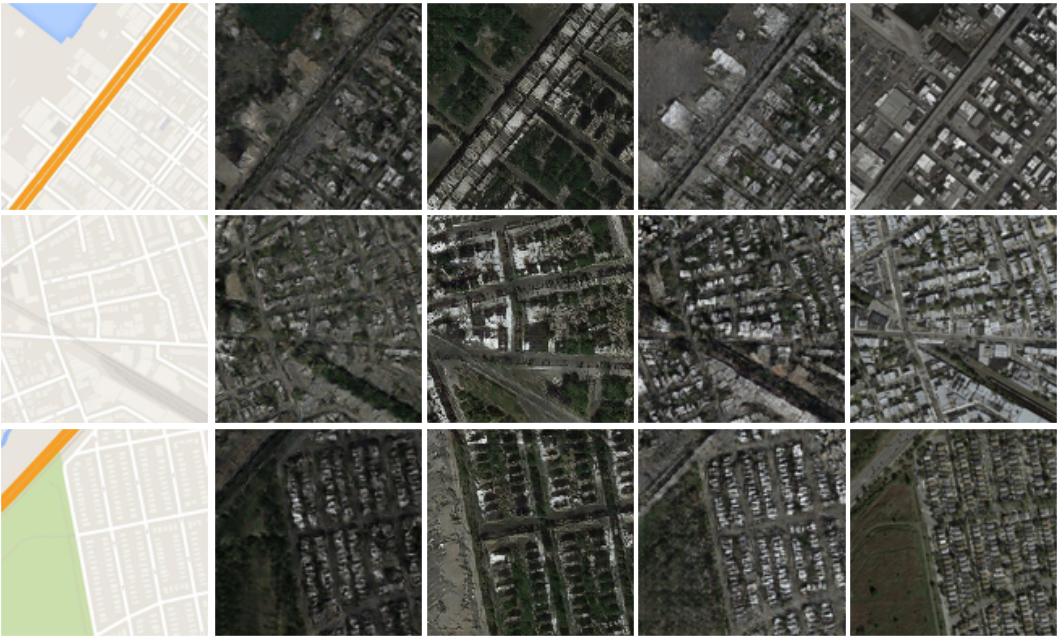}
\vspace{-0.2in}
\caption{Maps-to-satellite image conversion.}
\label{fig:conv5supp}
\end{figure}
\end{table}

\newpage
\begin{table}[H]
\centering
\vspace{0.15in}
\begin{tabu} to \textwidth {X[c] X[c] X[c] X[c] X[c]}
\large{\textbf{Original}} & \large{\textbf{CycleGAN}} & \large{\textbf{UNIT}} & \large{\textbf{InvAuto}} & \large{\textbf{Reference}}
\end{tabu}
\vspace{-0.3in}
\begin{figure}[H]
\centering
\includegraphics[width=\textwidth]{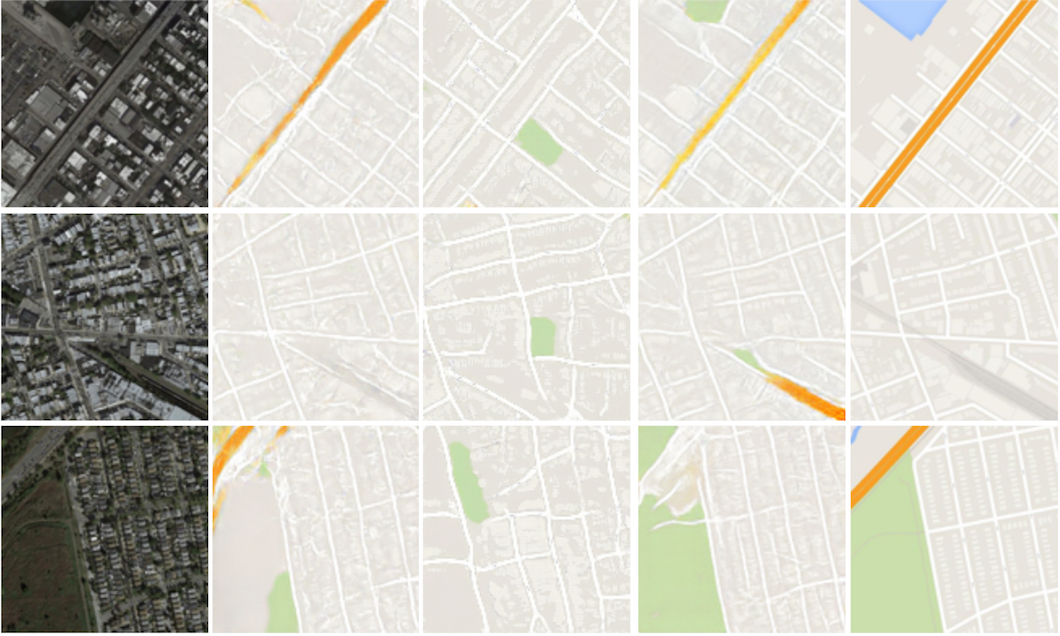}
\vspace{-0.2in}
\caption{Satellite-to-maps image conversion.}
\label{fig:conv6supp}
\end{figure}
\end{table}

\newpage

\begin{table}[H]
\centering
\vspace{0.15in}
\begin{tabu} to \textwidth {X[c] X[c] X[c]}
\textbf{Original} & \textbf{CycleGAN} & \textbf{InvAuto} 
\end{tabu}
\vspace{-0.3in}
\begin{figure}[H]
\includegraphics[width=\textwidth]{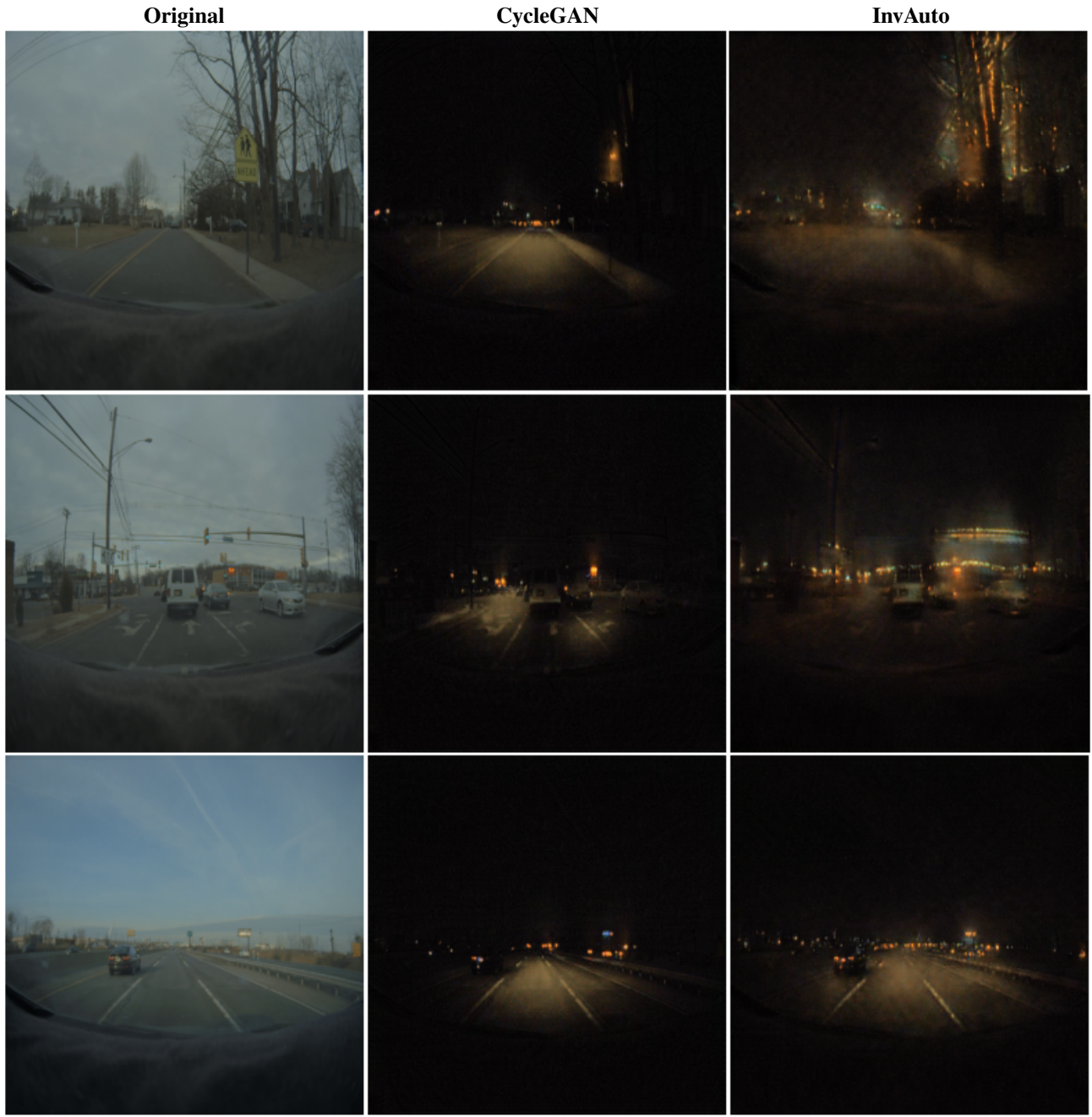}
\vspace{-0.25in}
  \caption{Experimental results with autonomous driving system: day-to-night conversion.} 
  \centering
  \label{fig:car1conv1}
\end{figure}
\end{table}

\newpage
\begin{table}[H]
\centering
\vspace{0.15in}
\begin{tabu} to \textwidth {X[c] X[c] X[c]}
\textbf{Original} & \textbf{CycleGAN} & \textbf{InvAuto} 
\end{tabu}
\vspace{-0.3in}
\begin{figure}[H]
\includegraphics[width=\textwidth]{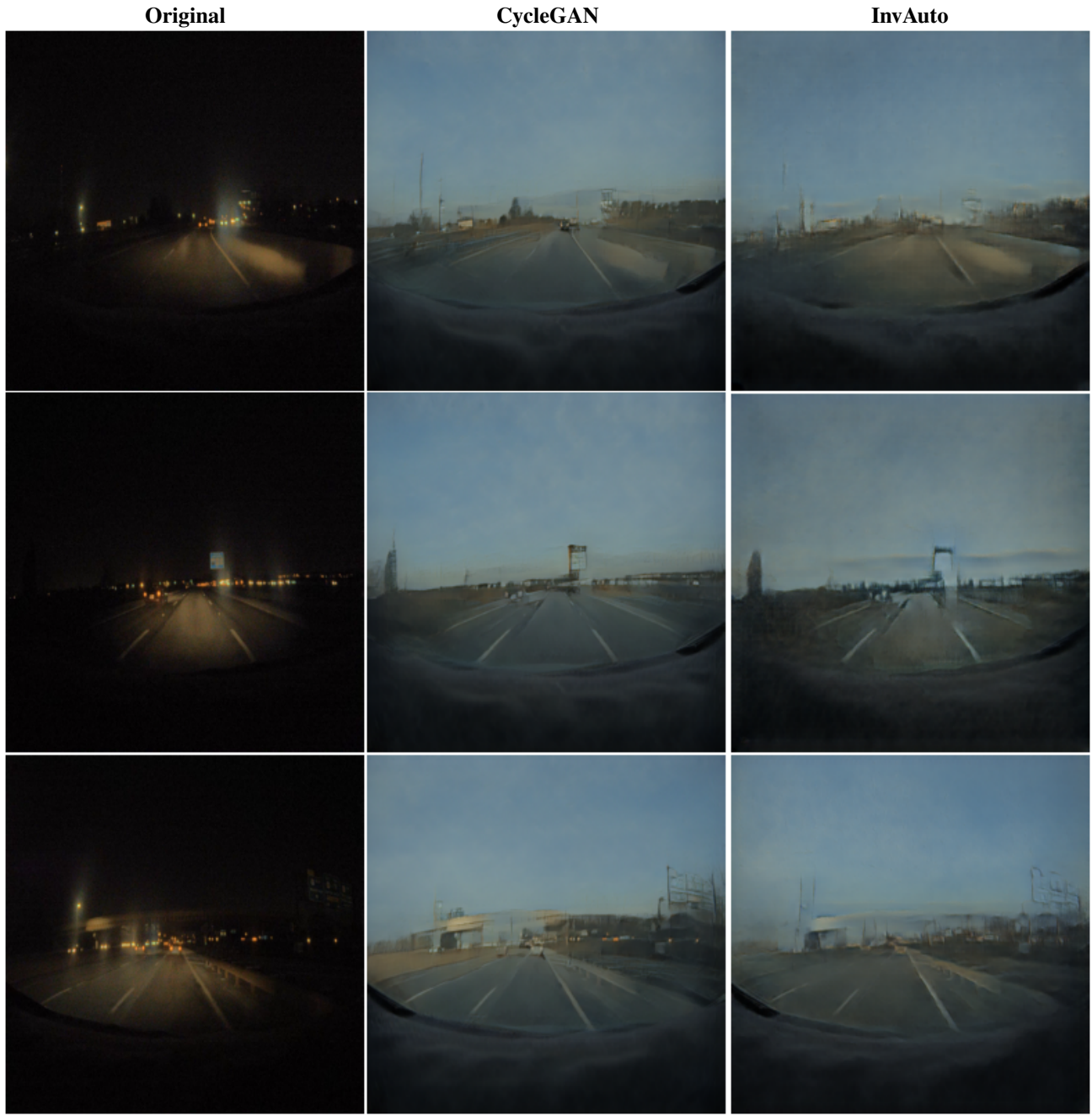}
\vspace{-0.2in}
\caption{Experimental results with autonomous driving system: night-to-day conversion.} 
  \centering
  \label{fig:car2conv2}
  \vspace{-0.1in}
\end{figure}
\end{table}

\newpage
\twocolumn
\section{Invertible autoencoder for domain adaptation: architecture and training}
\label{sec:C}

\textbf{Generator architecture} Our implementation of InvAuto contains $18$ invertible residual blocks for both $128 \times 128$ and $512 \times 512$ images, where $9$ blocks are used in the encoder and the remaining in the decoder. All layers in the decoder are the inverted versions of encoder's layers. We furthermore add two down-sampling layers and two up-sampling layers for the model trained on $128 \times 128$ images, and three down-sampling layers and three up-sampling layers for the model trained on $512 \times 512$ images. The details of the generator's architecture are listed in Table~\ref{gen_128} and Table~\ref{gen_512}. For convenience, we use Conv to denote convolutional layer, ConvNormReLU to denote Convolutional-InstanceNorm-LeakyReLU layer, InvRes to denote invertible residual block, and Tanh to denote hyperbolic tangent activation function. The negative slope of LeakyReLU function is set to $0.2$. All filters are square and we have the following notations: $K$ represents filter size and $F$ represents the number of output feature maps. The paddings are added correspondingly.

\textbf{Discriminator architecture} We use similar discriminator architecture as PatchGAN~\cite{pix2pix}. It is described in Table~\ref{dis}. We use this architecture for training both on $128 \times 128$ and $512 \times 512$ images.

\textbf{Criterion and Optimization} At training, we set $\lambda = 10$ and use $l_1$ loss for the cycle consistency in Equation~\ref{cycle_loss}. We use Adam optimizer\cite{kingma_ba_2017} with learning rate $l_r$ = 0.0002, $\beta_1 = 0.5$ and $\beta_2 = 0.999$. We also add $l_2$ penalty with weight $10^{-6}$.

\newpage

\begin{table}[H]
\centering
\begin{tabu}{||c |c| c||} 
 \hline
 Name & Stride & Filter  \\ [0.5ex] 
 \hline
 ConvNormReLU     &2 $\times$ 2& K4-F64\\ 
 \hline
 ConvNormReLU     &2 $\times$ 2& K4-F128\\
 \hline
 ConvNormReLU     &2 $\times$ 2& K4-F256\\
 \hline
 ConvNormReLU     &1 $\times$ 1& K4-F512\\
 \hline
 Conv             &1 $\times$ 1& K4-F1\\
 \hline
\end{tabu}
\caption{Discriminator for both 128 $\times$ 128 and 512 $\times$ 512 images.}
\label{dis}
\end{table}

\begin{table}[H]
\centering
 \begin{tabu}{||c|c|c||} 
 \hline
 Name & Stride & Filter  \\ [0.5ex] 
 \hline
 ConvNormReLU   & 1 $\times$ 1 & K7-F64\\ 
 \hline
 ConvNormReLU   & 2 $\times$ 2 & K3-F128\\
 \hline
 ConvNormReLU   & 2 $\times$ 2 & K3-F256\\
 \hline
 InvRes & 1 $\times$ 1 & K3-F256\\
 \hline
 InvRes & 1 $\times$ 1 & K3-F256\\
 \hline
 InvRes & 1 $\times$ 1 & K3-F256\\
 \hline
 InvRes & 1 $\times$ 1 & K3-F256\\
 \hline
 InvRes & 1 $\times$ 1 & K3-F256\\
 \hline
 InvRes & 1 $\times$ 1 & K3-F256\\
 \hline
 InvRes & 1 $\times$ 1 & K3-F256\\
 \hline
 InvRes & 1 $\times$ 1 & K3-F256\\
 \hline
 InvRes & 1 $\times$ 1 & K3-F256\\
 \hline
 ConvNormReLU   & 1/2 $\times$ 1/2 &K3-F128\\
 \hline
 ConvNormReLU   & 1/2 $\times$ 1/2 &K3-F64\\
 \hline
 Conv &1 $\times$ 1 & K7-F3  \\ 
 \hline
 Tanh       &               &   \\ 
 \hline
\end{tabu}
\caption{Generator for 128 $\times$ 128 images.}
\label{gen_128}
\end{table}

\begin{table}[H]
\centering
 \begin{tabu}{||c|c|c||} 
 \hline
 Name & Stride & Filter  \\ [0.5ex] 
 \hline
 ConvNormReLU   & 1 $\times$ 1 & K7-F64\\ 
 \hline
 ConvNormReLU   & 2 $\times$ 2 & K3-F128\\
 \hline
 ConvNormReLU   & 2 $\times$ 2 & K3-F256\\
 \hline
 ConvNormReLU   & 2 $\times$ 2 & K3-F512\\
 \hline
 InvRes & 1 $\times$ 1 & K3-F512\\
 \hline
 InvRes & 1 $\times$ 1 & K3-F512\\
 \hline
 InvRes & 1 $\times$ 1 & K3-F512\\
 \hline
 InvRes & 1 $\times$ 1 & K3-F512\\
 \hline
 InvRes & 1 $\times$ 1 & K3-F512\\
 \hline
 InvRes & 1 $\times$ 1 & K3-F512\\
 \hline
 InvRes & 1 $\times$ 1 & K3-F512\\
 \hline
 InvRes & 1 $\times$ 1 & K3-F512\\
 \hline
 InvRes & 1 $\times$ 1 & K3-F512\\
 \hline
 ConvNormReLU   &1/2 $\times$ 1/2 & K3-F256\\
 \hline
 ConvNormReLU   & 1/2 $\times$ 1/2 &K3-F128\\
 \hline
 ConvNormReLU   & 1/2 $\times$ 1/2 &K3-F64\\
 \hline
 Conv &1 $\times$ 1 & K7-F3  \\ 
 \hline
 Tanh      &             &   \\ 
 \hline
\end{tabu}
\caption{Generator for 512 $\times$ 512 images.}
\label{gen_512}
\end{table}

\end{document}